\documentclass[journal]{vgtc}                     % final (journal style)
\onlineid{1157}

\vgtccategory{Research}

\newcommand{\toolName}{{\textit{Conch}}}

\usepackage{cqhcmd}

\newcommand{\Aspiral}{\textbf{\textit{Archimedean spiral}}}
\newcommand{\Aspirals}{\textbf{\textit{Archimedean spirals}}}

\newcommand{\cds}{competitive debates}

\newcommand{\Cds}{Competitive debates}

\newcommand{\clashp}{\textbf{\textit{clash point}}}
\newcommand{\clashps}{\textbf{\textit{clash points}}}

\newcommand{\Clashp}{\textbf{\textit{Clash point}}}
\newcommand{\Clashps}{\textbf{\textit{Clash points}}}

\newcommand{\bk}{\textbf{\textit{block}}}
\newcommand{\bks}{\textbf{\textit{blocks}}}

\newcommand{\Bks}{\textbf{\textit{Blocks}}}

\newcommand{\PcsV}{Process View}
\newcommand{\StgV}{Strategy View}
\newcommand{\CttV}{Content View}
\newcommand{\SeshV}{Session View}

\usepackage{multirow}
\usepackage{longtable}
\usepackage{ltablex}
\usepackage{array}
\usepackage{booktabs}

\title{%
  \texorpdfstring{%
    \toolName{}: \underline{Co}mpetitive Debate A\underline{n}alysis via Visualizing \underline{C}lash Points and \underline{H}ierarchical Strategies
  }{%
    \toolName{}: Competitive Debate Analysis via Visualizing Clash Points and Hierarchical Strategies
  }%
}

\author{%
  \authororcid{Qianhe Chen}{0000-0001-8291-5789},
  \authororcid{Yong Wang}{0000-0002-0092-0793},
  \authororcid{Yixin Yu}{0009-0003-5193-8592},
  \authororcid{Xiyuan Zhu}{0009-0007-2059-2110}, 
  Xuerou Yu, 
  and \authororcid{Ran Wang}{0000-0002-4340-0018}
}

\authorfooter{
    \item
        Qianhe Chen, Y. Yu, X. Zhu, X. Yu and R. Wang are with School of Journalism and Information Communication, Huazhong University of Science and Technology (HUST).
        E-mail: qianhechen01@gmail.com.
     \item
  	Yong Wang is with the College of Computing and Data Science, Nanyang Technological University.
  	E-mail: yong-wang@ntu.edu.sg.
    \item 
        Ran Wang is the corresponding author, also with School of Future Technology, HUST.
        E-mail: rex\_wang@hust.edu.cn.
}

\abstract{
In-depth analysis of competitive debates is essential for participants to develop argumentative skills and refine strategies, and further improve their debating performance. 
However, manual analysis of unstructured and unlabeled textual records of debating is time-consuming and ineffective, as it is challenging to reconstruct contextual semantics and track logical connections from raw data. 
To address this, we propose \toolName{}, an interactive visualization system that systematically analyzes both what is debated and how it is debated. 
In particular, we propose a novel parallel spiral visualization that compactly traces the multidimensional evolution of clash points and participant interactions throughout debate process. 
In addition, we leverage large language models with well-designed prompts to automatically identify critical debate elements such as clash points, disagreements, viewpoints, and strategies, enabling participants to understand the debate context comprehensively. 
Finally, through two case studies on real-world debates and a carefully-designed user study, we demonstrate \toolName{}'s effectiveness and usability for competitive debate analysis. 
}


\teaser{
  \centering
  \includegraphics[width=\linewidth, alt={A view of a city with buildings peeking out of the clouds.}]{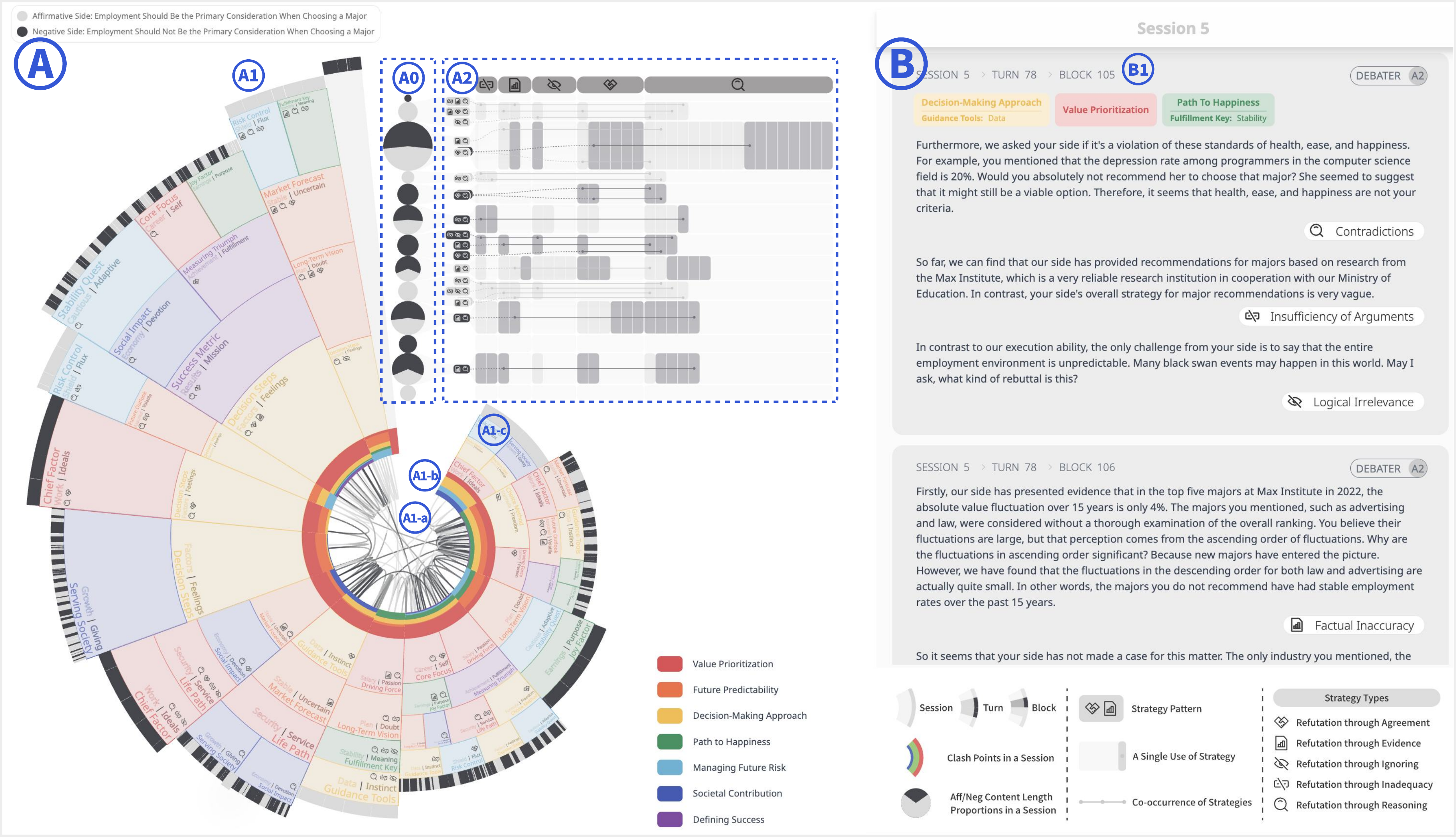}
  \caption{The user interface of \toolName{}. 
(A) Overview consists of two parts: the \PcsV{} (A0) and the \SeshV{} (A1) showing the evolution and interactions of debate among \bks{} based on \clashps{}, disagreements, and viewpoints; 
the Strategy View (A2) displaying the usage and co-occurrence of debate strategies. 
(B) The \CttV{} presents the specific textual content (B1), allowing users to closely examine arguments and strategies. 
In this figure, \toolName{} visually represents the evolution and interactions of debates among \bks{}, helping users intuitively understand how disagreements and strategies develop and interact over time.
  }
  \label{fig:teaser}
}

\graphicspath{{figs/}{figures/}{pictures/}{images/}{./}} % where to search for the images

\usepackage{tabu}                      % only used for the table example
\usepackage{booktabs}                  % only used for the table example
\usepackage{lipsum}                    % used to generate placeholder text
\usepackage{mwe}                       % used to generate placeholder figures

\usepackage{array}
\usepackage{graphicx}
\usepackage{makecell}
\usepackage{mathptmx}                  % use matching math font
\usepackage{amsmath}

\begin{document}

%%%%%%%%%%%%%%%%%%%%%%%%%%%%%%%%%%%%%%%%%%%%%%%%%%%%%%%%%%%%%%%%
%%%%%%%%%%%%%%%%%%%%%% START OF THE PAPER %%%%%%%%%%%%%%%%%%%%%%
%%%%%%%%%%%%%%%%%%%%%%%%%%%%%%%%%%%%%%%%%%%%%%%%%%%%%%%%%%%%%%%%

%% The ``\maketitle'' command must be the first command after the
%% ``\begin{document}'' command. It prepares and prints the title block.
%% the only exception to this rule is the \firstsection command
\firstsection{Introduction}

\maketitle

The competitive debate is a structured and competitive form of communication that challenges participants' comprehensive abilities, including logical thinking, expression skills, rapid analysis, argument construction, and rebuttal techniques \cite{zhang2024can}. By analyzing previous debates, they can learn effective strategies \cite{aryanti2024use}, identify common mistakes \cite{1sari2019error}, and understand how successful arguments are built \cite{4jes}. However, this process currently relies on manually reviewing long transcripts or videos \cite{3Supardi_Sayogie_2022}, which is time-consuming and makes it hard to track how arguments develop or connect across different parts of a debate. For example, a team might establish a strong argument early but fail to defend it later, and this common tactic detail often goes unnoticed during conventional manual analysis. Therefore, automated methods are essential to help debaters and coaches uncover hidden patterns and interaction dynamics in historical debates effectively and efficiently to improve performance.

Previous work on debate analysis (e.g., online \cite{12Bikakis2023Sketching}, formal \cite{3Supardi_Sayogie_2022}, political debates \cite{aryanti2024use}) has contributed to identifying argument components \cite{DELHOMME2022113694, 9Iman_Khazaali_2024}, labeling claims \cite{lawrence2020argument}, and applying predictive models to assess persuasiveness \cite{ruiz-dolz-etal-2023-automatic}. These approaches offer valuable insights into what is being said in a debate, especially at the sentence level \cite{16ALLS4636}. However, they remain insufficient to address two key challenges faced by debaters and coaches when analyzing \cds{}. On one hand, most existing methods focus on extracting claims, arguments, or keywords from individual sentences \cite{12Bikakis2023Sketching, 16ALLS4636, zhang2016conversational}. While this is useful for basic content analysis, it overlooks key elements that are essential to competitive debating, such as \clashps{}, disagreements, viewpoints, and strategies \cite{15athanasopoulos2023clash, bkedkowski2024systemic}. These elements usually connect multiple speaking turns and require examining the full debate context to properly identify and interpret them. 

On the other hand, current approaches pay little attention to how debates unfold over time. \Cds{} involve multiple sessions, speaker roles, and strategy interactions across different stages \cite{bkedkowski2024persuasive, zhang2024can}. Debaters often respond to earlier disagreements, adapt their strategies, and establish dominance at critical \clashps{} over the debate's progression \cite{bkedkowski2024systemic}. However, most existing methods for debate analysis typically process debates as disconnected claims or arguments \cite{lawrence2020argument}, overlooking how opposing views develop over time and interact across different debate stages to form meaningful argument structures \cite{12Bikakis2023Sketching, 6Lapesa_2020}. As a result, existing approaches struggle to adequately address two fundamental aspects of \cds{} for experts: what to debate and how to debate.

To address these challenges, we first introduce \clashps{} and refutation strategies as essential elements for analyzing \cds{}. 
\Clashps{}, representing the core disagreements between two sides, provide a focused understanding of critical conflict areas\cite{ericson2011debater}. 
Meanwhile, refutation strategies, such as evidence-based or reasoning-based refutation, illustrate how debaters systematically dismantle opposing arguments \cite{ericson2011debater, 2004DEBATING}. 
By combining these two elements, we can track the progression of debates more effectively, highlighting both the key points of contention and the strategy approaches used by debaters and coaches. 
This approach allows for a more detailed and dynamic analysis of debate interactions, capturing both the structure and tactics central to \cds{}.

We propose \toolName{}, an interactive visualization system that enhances the analysis of debate competitions by summarizing debate dynamics and visualizing key logical interactions. 
The system consists of two primary views: an overview and a detail view. The overview is divided into two parts: 
(1) A process view, which illustrates the evolution of \clashps{}, their progression, and the interactions between debaters over time; 
(2) A strategy view, which depicts the distribution and co-occurrence of various refutation strategies used throughout the debate. 
These visualizations provide users with a comprehensive and in-depth understanding of both the argumentative structure and strategy choices. 
Additionally, the detail view complements the overview by displaying the specific textual content of the debate, allowing users to closely examine the arguments and strategies presented. 
The spiral-shaped design of \toolName{}, inspired by the natural growth patterns of conch shells, uses circular timelines to show how \cds{} develop and interact over time.
The effectiveness of \toolName{} was demonstrated through two case studies and a carefully-designed user interview. 
\toolName{} can be publicly accessed\footnote{\href{https://debate.datavizu.app/}{https://debate.datavizu.app/}}. 
In summary, our main contributions are:

\begin{enumerate}
    \item An interactive system \toolName{}, for the first time, is proposed to facilitate the analysis of \clashps{} and strategy interactions hierarchically in \cds{} for debaters and coaches.
    \item A compact parallel spiral visualization is designed to represent the temporal and structural evolution of \cds{} centered on \clashps{}. Inspired by \Aspirals{}, the layout is arranged to optimize space while enabling block-wise exploration of debate dynamics across sessions.
    \item An augmented stacked bar chart design that enables detailed comparison of strategies both within and across different debate sessions, providing a comprehensive view of strategy patterns.
    \item Two case studies and a user study are conducted on two distinct datasets with experts on \cds{} to show the usefulness and effectiveness of our system.
\end{enumerate}

\section{Related Work}

\textbf{Argument Analysis.} 
Research in argument visualization has focused on representing the logical structure of arguments. 
These works use diagrams and graphs to map out the elements of an argument, such as premises and conclusions, and the relationships between them \cite{Zhou_2019, Cullen_2018}. 
The primary goal is to help users understand complex reasoning within static texts \cite{Khartabil2021}. 
However, these methods are designed for stable content where arguments are already complete and presented in a sequential order. 
They are not well-suited for live debate competitions, which are dynamic and involve real-time, back-and-forth exchanges \cite{2004DEBATING}. 
When we look specifically at the analysis of debate competitions, existing studies tend to focus on the debate process. 
They examine aspects like how debaters construct their arguments, how they interact, and the rhetorical strategies they employ \cite{26AgustiniPutri_2023, 11Freddie2020Learning, 16ALLS4636}. 
While this research provides valuable insights into the components of a debate, it often analyzes arguments or strategies in isolation. 
It gives less direct attention to \clashps{} — the specific moments where opposing arguments directly confront each other — and the detailed refutation strategies debaters use in response. 
Therefore, our work addresses this gap by introducing a framework for systematically identifying and categorizing \clashps{} and the refutation strategies. 
This allows us to provide a deeper understanding of the strategic core of competitive debating.

\textbf{Visual Analytics of Presentation.} 
While some research focuses on visualizing single-presenter performance \cite{34Zeng_2019, 33Wang_2020}, our work is concerned with visualizing the interactions among multiple participants, which is essential for understanding debates. 
Existing approaches in this area typically visualize two main aspects: discussion content and participant interactions. 
For the discussion content, researchers often extract high-level topics using methods like Latent Dirichlet allocation (LDA) 
or word embeddings to show what is being discussed \cite{9331282, chen-etal-2016-wordforce}. 
For interactions, they commonly use network graphs or timelines to show connections between speakers or the sequence of turns \cite{seki2020visualization, celepkolu2021designing}. 
However, these established methods have two key limitations in the context of a debate. 
First, topic-level views are often too general; they show what is being discussed but lose the specific arguments and the logical connections between them \cite{dang2018commodeler}. 
Second, common interaction visualizations show that participants are interacting but not how. 
For example, a simple line connecting two speakers cannot distinguish a basic reply from a direct, strategic refutation \cite{9331282}. 
Our work is designed to overcome these limitations. 
To solve the problem of over-generalization, our system visualizes the debate using a layered structure of \clashps{}, disagreements, and viewpoints, which preserves the essential argumentative context. 
To reveal the nature of interactions, our novel visual design explicitly shows the temporal flow and strategic purpose of each exchange. 
This allows users to see not just that a clash occurred, but precisely how an argument was constructed, challenged, and refuted over time.

\textbf{Text Visualization.} 
Text visualization can be broadly classified into node-link diagrams, matrix-based layouts, spatial region encodings, and glyph-augmented representations \cite{7156366, computers8010017}. 
Node-link diagrams are dominant for visualizing hierarchical and relational data \cite{https://doi.org/10.1111/cgf.12378, 9716779, 9552910}. 
Their applications include tree and graph layouts that preserve structural links \cite{10.2312:eurovisshort.20171153, 6816054}, as well as enhanced versions like extended Chord diagrams that reveal keyword co-occurrence patterns in social media \cite{10.2312:eurovisshort.20171149}.
Matrix-based techniques facilitate pattern detection in sequential discourse by representing similarity across utterance pairs \cite{6400853, https://doi.org/10.1111/cgf.13179, 10.1145/3132169}. 
Spatial layouts present a radial diagram for script-based stories \cite{Watson_2019}, using concentric rings to encode scene settings, character presence, and emotional arcs, enabling structural and comparative analyses via difference overlays. 
Glyph-based forms integrate micro-visualizations like sparklines into tag clouds\cite{5613457} or node-link diagrams\cite{7185421} to simultaneously convey temporal trends and categorical frequency. 
These diverse design strategies demonstrate how structural properties of text — such as hierarchy, temporal sequence, or faceted annotation — can be mapped to distinct visual encodings, with each encoding supporting specific tasks such as exploration, comparison, or trend analysis. 
However, such layouts struggle with the unique structure of competitive debates: long-range argument threads spanning sessions create visual clutter in linear or tree representations.
Furthermore, they typically visualize general topics or sentiment. 
None of them can capture the deeper, domain-specific semantic hierarchy of \clashps{}, disagreements, and refutation strategies, which is critical for debate analysis and is the focus of this work.

\section{Requirements Analysis}
\label{req}
To better understand the major challenges and design requirements of debate competition visualization, we worked closely with six debaters (\debater{1,2,3,4,5,6}), four debate coaches (\coach{1,2,3,4}), and three experts (\textbf{E1-3}) for over 6 months. 
We used snowball sampling \cite{Snowball} to find debaters and coaches with experiences of debate competitions across different languages (Chinese and English) and multiple formats with a wide range of experience (1 - 6 years). 
We conducted a series of semi-structured interviews with them; each interview lasted for more than one hour. 
Details regarding each participant and interview questions are available in the Supplementary Materials. 
Following their feedback, we summarized major design requirements into \textbf{R1-R4} as follows:

\begin{enumerate}[label=\textbf{R{\arabic*}}, nolistsep]
\item 
\textbf{Explore the overall debate evolution across sessions.} 
Post-competition analysis is a crucial practice for debaters and coaches, as demonstrated in our interviews. 
They review debates to capture the structure and evolution of debates. 
\debater{1}, \debater{4}, and \debater{5} usually summarized each session separately, while \debater{3} and \coach{3} also highlighted that different sessions have specific functions. 
However, they found it was time-intensive and required additional effort to trace the overall evolution. 
Therefore, it is crucial to present debate's chronological evolution across sessions, enabling them to quickly understand the debate's development.

\item 
\textbf{Identify and analyze clash points, disagreements, viewpoints, and strategies.}
Compared to other types of debate, \cds{} often involve more direct and explicit conflicts between debaters. 
Therefore, clearly identifying and analyzing these specific debate elements is particularly important. 
However, debaters and coaches found it difficult to accurately capture these elements from recordings and recall them later. 
Specifically, \debater{2} mentioned that understanding \clashps{} and disagreements helped identify main conflicts and key arguments. 
Both \debater{5} and \coach{3} emphasized that identifying debaters' viewpoints on disagreements helped understand core differences between two sides. 
Moreover, all coaches agreed that identifying the debater strategies, such as refutation and defense approaches, was essential for analyzing performance and improving training. 
Therefore, it is necessary to clearly visualize these debate elements and allow users to explore their interrelationships.

\item 
\textbf{Analyze the interaction within each disagreement.} 
Analyzing how debaters interact within each disagreement is crucial for effective debate analysis, as emphasized by debaters and coaches. 
In particular, \debater{4} and \coach{2} mentioned that examining how debaters refute one another revealed the logic and reasoning of their arguments. 
Moreover, tracking the interactions between opposing sides helped understand how each disagreement evolves. 
Such analysis helps debaters and coaches to assess the strengths and weaknesses of each side's argumentation. 
Therefore, a clear and intuitive visualization of these interactions would be highly valuable for debaters and coaches.

\item 
\textbf{Examine the detail of original text content.} 
Beyond general overviews, debaters emphasized the need to analyze and learn from the concrete language used in debates. 
\coach{1} noted that the text was useful for learning strategy applications directly, and \coach{2} highlighted it as a resource for in-depth study. 
Therefore, it is necessary to present the detailed text, including specific strategies and the corresponding language.
\end{enumerate}

\section{Data Abstraction and Processing}
\label{data}

As shown in \cref{tab:dataset}, we collected a dataset from three representative competitive debates: the top-tier Chinese debate competition - International Chinese Debate Invitational\footnote{\href{https://bilibili.com/video/BV1Ca4y1y72t/}{https://bilibili.com/video/BV1Ca4y1y72t/}} (ICDI), the largest academic competition - National Speech and Debate Tournament\footnote{\href{https://youtu.be/ytA7V3wXux0/}{https://youtu.be/ytA7V3wXux0/}} (NSDT), and the most prestigious debating society - the Oxford Union\footnote{\href{https://youtu.be/ByrhYvbwJoA/}{https://youtu.be/ByrhYvbwJoA/}} (TOU). 
This collection includes debates conducted in two languages (Chinese and English) and three formats, offering a cross-cultural perspective on debate structures and strategies. 
Notably, interactions are more frequent and direct in ICDI than other competitions.

\begin{table}[htbp]
\centering
\caption{Overview of the debate competition dataset}
\label{tab:dataset}
\resizebox{\columnwidth}{!}{%
\begin{tabular}{ccccccc}
\toprule
\textbf{Competition} & \textbf{Language} & \textbf{Debater} & \textbf{Session} & \textbf{Turn} & \textbf{Duration} & \textbf{Word} \\ \midrule
ICDI                 & Chinese               & 8                & 13             & 181            & 1.1 h             & \textasciitilde 12 K          \\
NSDT                 & English               & 6                & 8              & 8              & 1 h               & \textasciitilde 10 K          \\
TOU                  & English               & 8                & 8              & 8              & 1.5 h             & \textasciitilde 12 K          \\ \bottomrule
\end{tabular}%
}
\end{table}

\begin{figure}
    \centering
    \includegraphics[width=\linewidth]{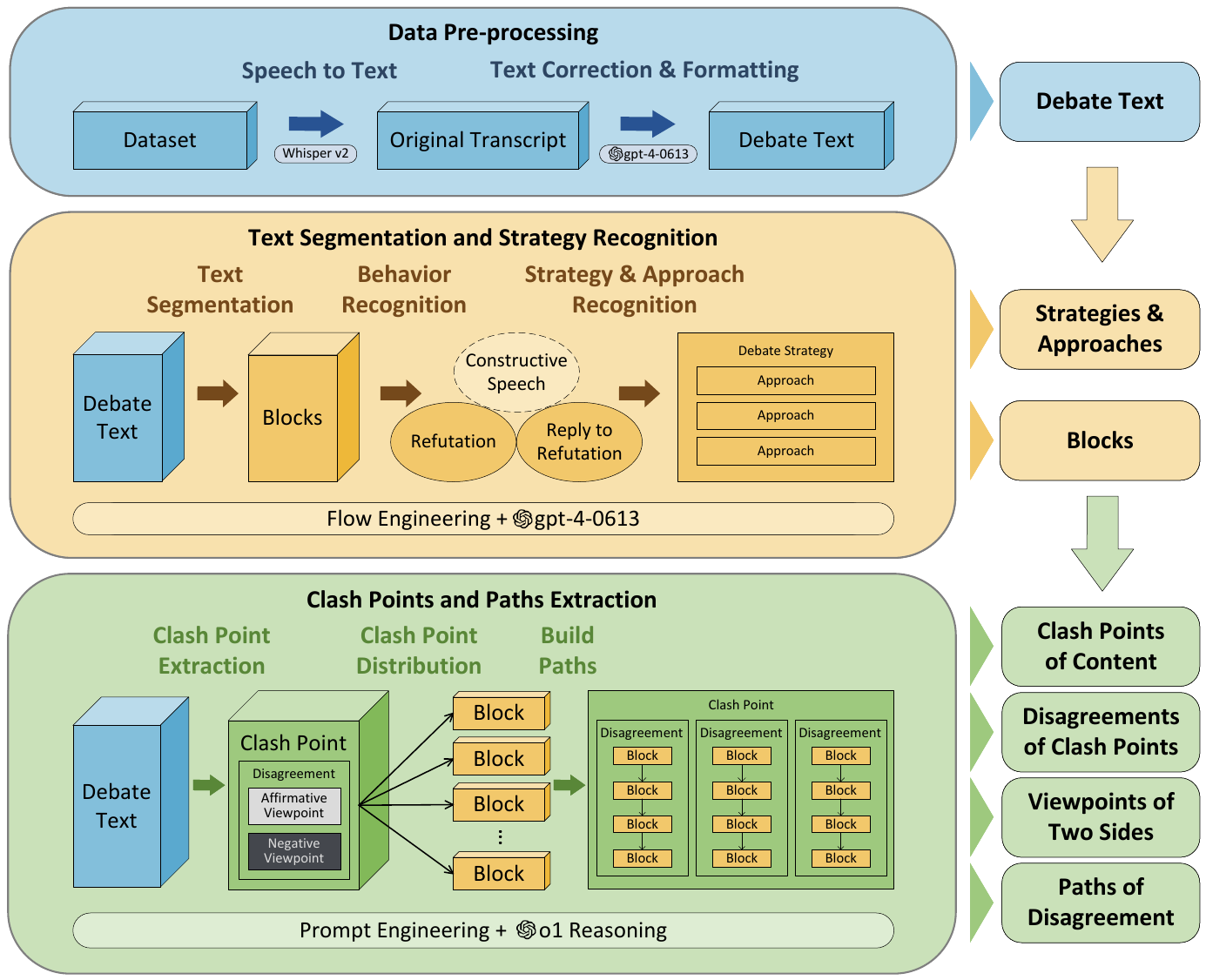}
    \caption{The technical framework for segmenting \bks{}, extracting \clashps{}, disagreements, and viewpoints, and constructing interaction paths.}
    \label{fig:tech-flow}
\end{figure}

\textbf{Debate text extraction and structuring.} 
The debate text was first extracted from debate videos and then structurally divided. 
A debate has a clear and well-defined structure. 
It consists of \textbf{sessions}, \textbf{turns}, and \textbf{\bks{}}. 
\textbf{Sessions} are parts of the debate defined by the debate rules. \textbf{Turns} are uninterrupted speeches given by a single debater within a session. 
\Bks{} are the smallest units of debate content, defined as brief arguments that include reasoning and evidence \cite{ericson2011debater}. 
A session is consisted of one to several turns, and a turn is consisted of one to several \bks{}. 
Firstly, we transcribed the audio content using Whisper \cite{radford2022robustspeechrecognitionlargescale}, and used the GPT-4 \cite{openai2024gpt4technicalreport} to automatically correct errors in the initial transcription\footnote{\href{https://platform.openai.com/docs/guides/speech-to-text/}{https://platform.openai.com/docs/guides/speech-to-text/}}. 
Based on the transcripts and associated competition information, we annotated the debate sides, debaters,  sessions and turns of the textual content. 
To further evaluate the debate structure, we used large language model (LLM) techniques and designed prompts to guide GPT-4\footnote{The exact model of GPT-4 used in this paper is gpt-4-0613.} in segmenting debate texts into more integrated \bks{}. 
The segmentation results underwent manual review by three debate experts, achieving an average precision of 93.2\% with a Fleiss' Kappa score \cite{Fleiss1971} of 0.89, thereby validating the robustness of this approach.

\textbf{Construction of hierarchical debate analysis framework.} 
We constructed a hierarchical debate analysis framework by combining the strategy framework and the content framework. 
% For the analysis of debate strategies and approaches, a framework for organizing refutations in debate competitions is firstly established by integrating instructional books on debating \cite{ericson2011debater, 2004DEBATING, quinn2009debating}, as shown in \cref{tab:strategy}. 
To analyze debate strategies and approaches, a framework for organizing refutations in debate competitions was first established by integrating instructional books on debates \cite{ericson2011debater, 2004DEBATING, quinn2009debating}, which is provided in the Supplemental Material.
This framework provides a comprehensive coverage of refutation strategies in debates. 
Strategies and approaches of \bks{} were identified by GPT-4. The performance of GPT-4 was enhanced through a series of detailed prompts utilizing LangGPT \cite{wang2024langgpt}, few-shot learning \cite{min-etal-2022-rethinking}, and flow engineering \cite{ridnik2024codegenerationalphacodiumprompt}. 
Prompts included at least three standardized examples for each strategy facilitated few-shot learning. 
All identification results were manually validated by three debate experts and achieved an average precision of 95.2\% with a Fleiss' Kappa score of 0.95, validating the accuracy and reliability of our method. 

Additionally, we introduced a content framework to analyze debate content by first identifying \textbf{\clashps{}}, then examining \textbf{disagreements}, and finally extracting \textbf{viewpoints}.
\Clashp{} is a high-level concept, which is defined as the fundamental disagreement between the two sides and all the other disagreements simply derive from that first disagreement. 
Disagreements derive from \clashps{}, and viewpoints refer to the different opinions each side holds in response to these disagreements. 
The development process of each disagreement can be summarized as a \textbf{path}, which is a sequence of connected \bks{}.

The identification of \clashps{}, disagreements, viewpoints and paths requires understanding of long context and complex logic, which is challenging to be accomplished in a single step. 
As a result, we designed a three-step approach to leverage the strongest capabilities of reasoning LLMs at each stage, ultimately ensuring accurate results.  
First, we used the OpenAI o1 \cite{openai2024openaio1card} to precisely extract all \clashps{} and related disagreements with affirmative and negative viewpoints from the entire debate text, using the \clashp{} and disagreement definition as an instruction. 
Each \clashp{} is a phrase consisting of 2 to 4 words, and each disagreement is a phrase with 2 or 3 words, while viewpoints are concluded into single words. 
Then, \clashps{} and disagreements were distributed to \bks{} where they were referenced via prompts and gpt-o1\footnote{The exact model of o1 used in this paper is o1-2024-12-17 with high reasoning effort.} . 
\Bks{} with less than 20 words are considered too short to contain \clashps{} and disagreements. 
Each \clashp{} and disagreement may be distributed across multiple \bks{} and each \bk{} may be associated to multiple \clashps{} and disagreements. 
The distribution of \clashps{} and disagreements was manually validated by three debate experts using the same method, achieving an average precision of 92.7\% with a Fleiss' Kappa score of 0.95. 
To thoroughly analyze the relationship between these \bks{}, specially designed prompts and o1 were used to identify \bks{} with logical relevance. 
Mutually related \bks{} can be connected into relational paths, and each disagreement is represented by a path consisting of multiple \bks{}. 
All paths have been verified by three experts and are considered to meet their professional analysis needs, with essentially no omissions in terms of quantity.

\textbf{Expert Validation Methodology.} 
To validate our LLM-based method, we recruited three experts, each with over five years of debate experience, to create a ground truth by annotating our dataset. 
The annotation process followed a structured protocol to ensure consistency \cite{Braun2024}. 
First, we conducted a training session to familiarize experts with the annotation tasks. 
Next, they performed a pilot annotation on a small data sample and discussed their results to establish a unified annotation standard. 
After reaching an agreement on the standard, they proceeded with the formal annotation. 
During this process, any disagreements were resolved by majority rule. 
In the rare case that all three experts had different annotations, they would discuss the case together to determine the final label. 
Based on these annotations, we calculated the inter-annotator agreement using Fleiss' Kappa \cite{Fleiss1971}. 
The resulting high level of agreement confirmed the reliability of our ground truth. 
The specific technical workflow and all prompts are available in the Supplemental Material.

\section{Visual Design}
The visual interface of \toolName{} supports the exploration of debate competition evolution both in \clashps{} and in strategies with intuitive visualization designs. 
\cref{fig:teaser} shows the snapshot of the interactive system annotated with an overview (including the \SeshV{} (A0), the \PcsV{} (A1), and the \StgV{} (A2)), and the \CttV{} (B). 
The \PcsV{} (\cref{fig:teaser}A1) provides an integrated visualization of the temporal evolution of the debate across sessions, the main \clashps{} and the interactions among \bks{}, 
and the \StgV{} (\cref{fig:teaser}A2) reveals the co-occurring strategies and their usages in each session. 
The \SeshV{} (\cref{fig:teaser}A0) shows all sessions in the competition, and connects the \PcsV{} and the \StgV{}, 
while the \CttV{} (\cref{fig:teaser}B) offers detailed insights into the specific content. 
\toolName{} also facilitates interactive and collaborative exploration across the three views. 
A unified color encoding scheme is used throughout the interface to distinguish between the affirmative (white) and negative (black) sides and differentiate \clashps{} through distinct colors.

\subsection{\PcsV{}}
The \PcsV{} (\cref{fig:teaser}A1) is designed to demonstrate the evolution of debate (\textbf{R1}), highlighting \clashp{} and disagreements (\textbf{R2}), and their interactions across sessions (\textbf{R3}). 
It consists of a novel designed conch-like diagram outside (\cref{fig:teaser}A1-c), a chordal graph inside (\cref{fig:teaser}A1-a) and a ring of color-filled blocks in the middle (\cref{fig:teaser}A1-b). 
These elements collaboratively present an overview of the debate's evolution and interactions among \bks{} based on \clashps{}.

\begin{figure}[htbp]
    \centering
    \includegraphics[width=\linewidth]{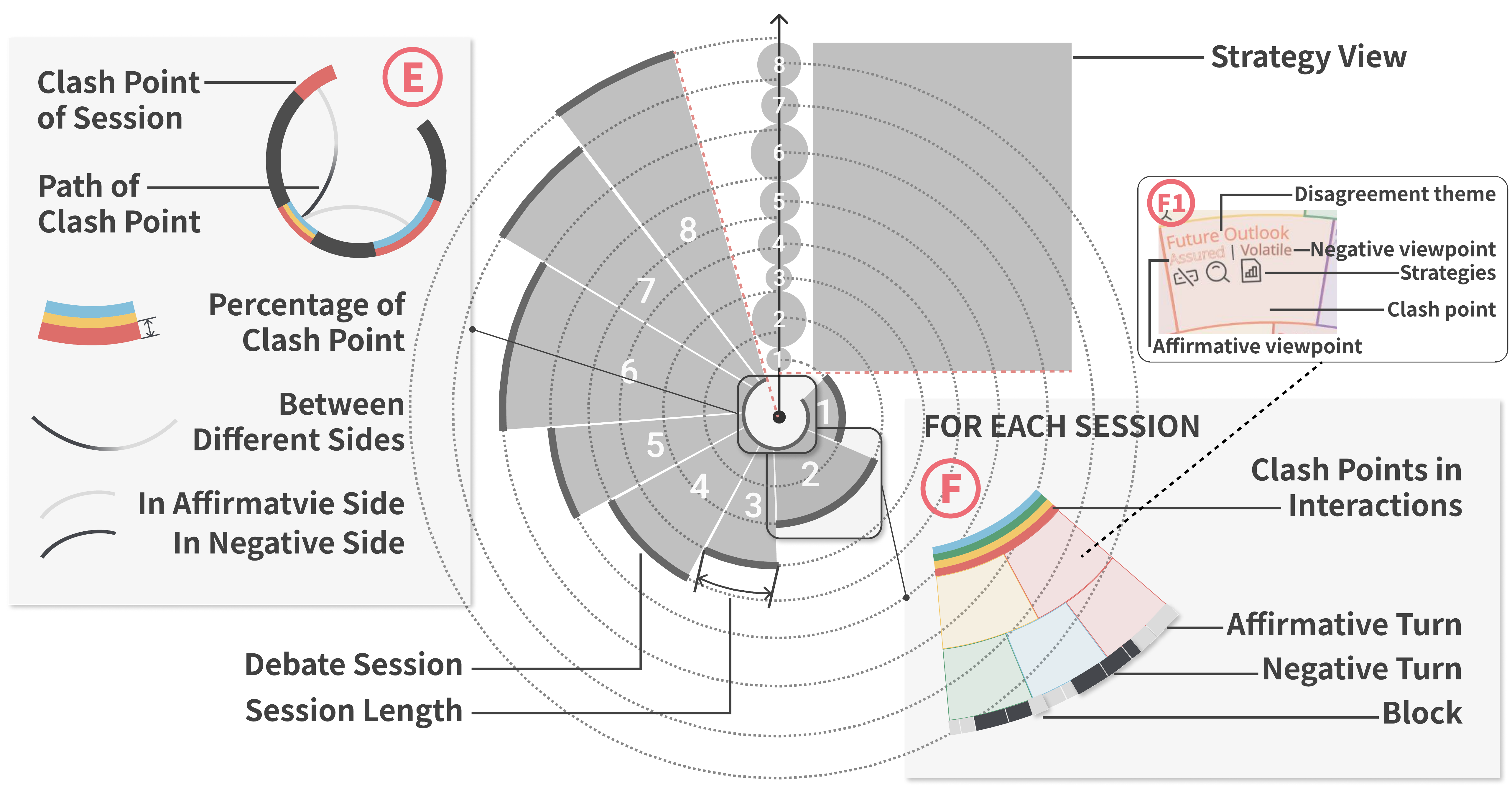}
    \caption{
The glyph design of the Process View. (E) illustrates the chord diagram and its surrounding elements; (F) illustrates the structure of the sector-shaped areas; (F1) illustrates the disagreement block content.
}
    \label{fig:a-spirals}
\end{figure}

The conch-like diagram (\cref{fig:teaser}A1-a) is composed of a series of parallel \Aspirals{} segments. 
As shown in \cref{fig:a-spirals}, each \Aspiral{} segment corresponds to a debate session, arranged clockwise in chronological order, with its arc length representing the session's content length. 
Each white and black segment (\cref{fig:a-spirals}F) within the \Aspiral{} represents a \bk{}, with its arc length indicating the content length and color encoding two sides. 
With all \Aspirals{} constructed in a vertically oriented polar coordinate system, their start points are positioned on the axis and serve as centers for a column of tangent circles (numbered 1 - 8 in \cref{fig:a-spirals}) in the \SeshV{}. 
These circles connect the \PcsV{} and the \StgV{}, which are designed to represent the session's content length and two sides' content proportions.

To analyze the \textbf{\textit{clash-points}}-based interactions among \bks{} (\textbf{R3}), a chordal diagram is integrated inside the \Aspirals{} (\cref{fig:teaser}A1-a), where each filled chord illustrates an interaction over a \clashp{} between two \bks{} by connecting their respective sessions. 
As shown in \cref{fig:a-spirals}E, the start and end points of the chord on each session's arc indicate the \bk{}'s position within that session. 
The color of the chord indicates whether the two connected parties are on the same side or different sides regarding the \clashp{}. 
If the parties are from different sides for this \clashp{}, the line will display two distinct colors. 
Each color occupies half of the line and corresponds to the color of one party's side. 
Conversely, if both parties belong to the same side concerning this \clashp{}, the entire line will be a single color, representing their side's color. 
Users are allowed to filter chords over a certain \clashp{} or \bk{}. 
The periphery of the chordal graph is an \Aspiral{} (\cref{fig:teaser}A1-b and \cref{fig:teaser}A1-c). 
As shown in \cref{fig:a-spirals}F, the area near the pole (\cref{fig:teaser}A1-b) is radially divided into multiple color-filled sections. Each section represents a \clashp{} in this session, and its width indicates the percentage of its corresponding \clashp{}. 
The remaining central spiral and its adjacent sector-like areas (\cref{fig:teaser}A1-c) display detailed information about sessions (\textbf{R2}). 
Each sector-like area (\cref{fig:a-spirals}F), representing a single session, is further divided into multiple blocks by the disagreements within the session. 
The area of each block is proportional to the number of \bks{} in the corresponding disagreement, while their order is not meaningful. 
Each block (\cref{fig:a-spirals}F1) is then filled with a specific color to indicate the \clashp{} under discussion. 
In each block, the theme of the disagreement is highlighted in a larger font, while the viewpoints of the opposing sides are juxtaposed in a smaller font. 
Additionally, the debate strategies used in the disagreement are depicted as icons in the innermost layer of the area. 
All content is organized into blocks and arranged along a curve. 
The font size in each block is maximized based on the available space and the block's proportional size. 
While this approach makes efficient use of space, it can cause some reading difficulty \cite{10.1145/2858036.2858057}. 
To address this, we provide a popup that displays the full text content when the user hovers over a block, ensuring better readability. 
Notably, a debater might mention a \clashp{} in a session, but it does not become part of an interaction with another debater. 
In these situations, the color for that \clashp{} will appear in the middle ring (\cref{fig:teaser}A1-b) to show the topic was present, but there will be no corresponding disagreement block in the outer diagram (\cref{fig:teaser}A1-c). 
This reflects the reality of debates, where not every point raised leads to a direct conflict.

\textbf{Design Alternatives and Advantages.} 
A key design requirement is to show the entire multi-session debate on a single screen. 
This allows easy analysis of chronological evolution across sessions (\textbf{R1}) and the interactions among disagreements (\textbf{R3}). 
At the same time, users must be able to see specific details like disagreements, viewpoints, and debate strategies (\textbf{R2}). 
These goals created two constraints for our \PcsV{}. 
First, the view needs enough internal space to present the debate's step-by-step progression alongside detailed information. 
Second, it must work well with external views, such as the \StgV{}. 
% These needs naturally pointed to a timeline-based visualization.
While alternative approaches such as multiple coordinated views could also be considered, we adopt a unified layout due to the tight coupling between temporal structure, debating content, and interaction strategies in a single page. 
Thus, a timeline-based visualization was proposed to balance overview and detail within a single integrated view.

To select a suitable layout for debate visualization, we compared three common timeline designs: linear, radial, and spiral \cite{Di_Bartolomeo_2020, 7581076}. 
Linear layouts are intuitive for showing chronological order \cite{El_Assady_2018, Cantareira_2024}, but struggle with long-range interactions common in debates (e.g., when a speaker in the final session rebuts a point from the first). 
Visualizing these connections creates long, overlapping lines, which leads to visual clutter and difficulties in tracing argument flow in a single screen. 
Radial layouts can display an entire debate on a single screen \cite{wang2022rumorlens, 7581076}. 
However, their closed structure makes it difficult to integrate with other information views in a coordinated display. 
Therefore, we chose a spiral layout. 
Its open and extensible nature is ideal for connecting the timeline with other views. 
Building on prior work confirming the spiral's space efficiency \cite{963273, 10.1145/288392.288399}, we developed a multi-spiral design. 
This approach provides more sufficient and adjustable space for visualizing the temporal progression and related details in a compact way.
 
Our multiple \Aspirals{} design employs a multi-spiral layout to represent debate sessions, but its primary challenge is achieving visual balance. 
In our approach, concentric spirals expand outwards from the center, with each spiral segment corresponding to a single session. 
This arrangement naturally shows the debate's temporal progression. 
However, since session's content have variable lengths, a simple layout would make some segments too large while others become illegibly small. 
To address these challenges, we propose two key solutions: an adaptive layout algorithm and a separate view for interaction links. 
First, our novel algorithm balances the layout. 
It maps each session's content to a specific arc length, then optimizes the radius and central angle for all segments simultaneously. 
This process minimizes space and prevents extreme size differences.
Second, to reduce visual clutter, we move all interaction links to a central chord diagram. 
Links connecting the diagram to the spiral are shown only on-demand when a user selects the related \clashp{}, keeping the main timeline clean. 
This integrated visualization is compact, balanced, and easy to interpret. 
Its adaptability makes the design generalizable for different debates, fulfilling the criteria \cite{7581076} for a viable design: being purposeful, interpretable, and generalizable. 
A detailed explanation of our algorithm is available in the Supplemental Material.

\subsection{\StgV{}}
\StgV{} (\cref{fig:teaser}A2) is designed to simultaneously show the usage of debate strategies and any strategies that appear together (\textbf{R2}). 
An augmented stacked bar diagram is used to help compare how often strategies are used by opposing sides. To support easy exploration, co-occurring strategies are positioned between the \SeshV{} (\cref{fig:teaser}A0) and the relevant augmented stacked bar for each session.

\begin{figure}[htbp]
  \centering
  \begin{subfigure}[t]{0.48\linewidth}
    \centering
    \includegraphics[width=\linewidth]{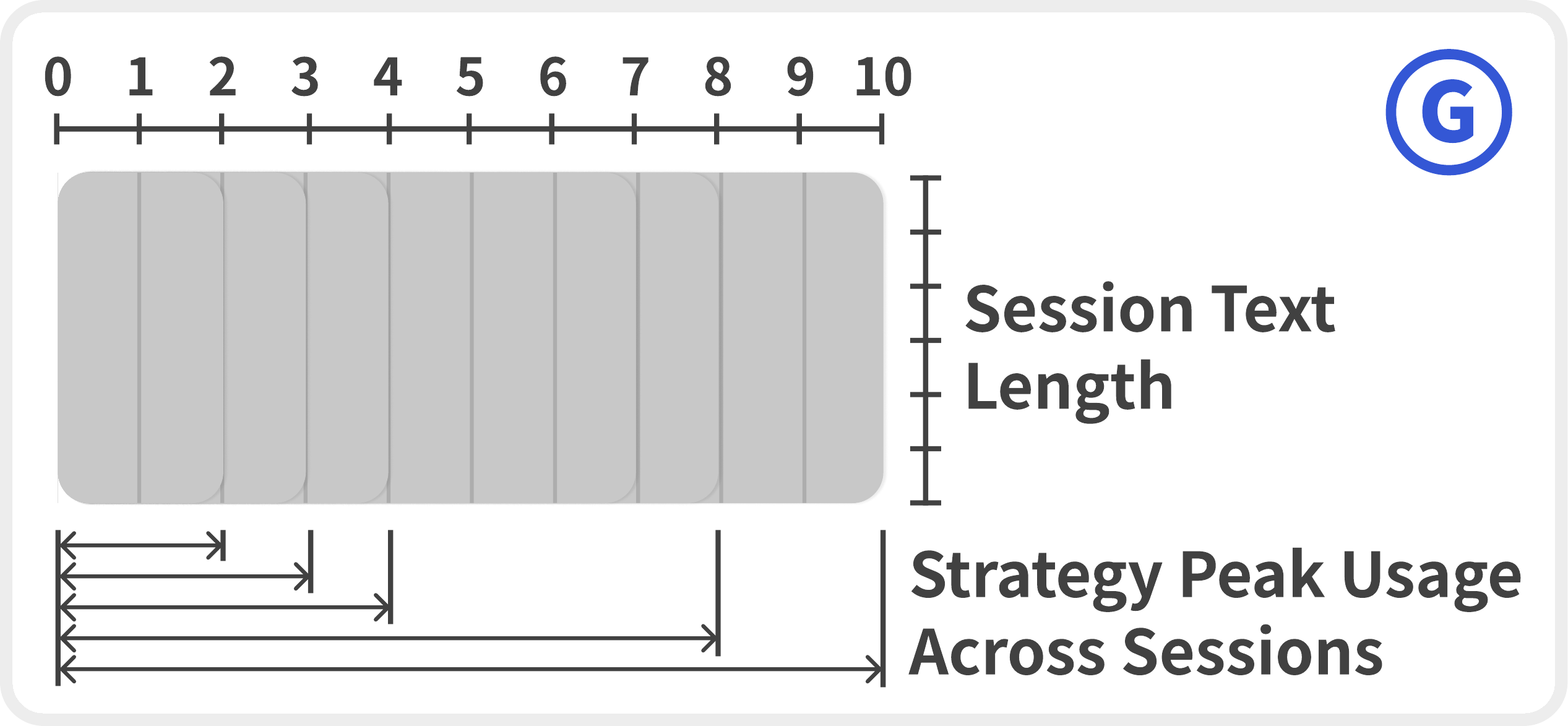}
    % \caption{}
    % \label{fig:sub1}
  \end{subfigure}
  \hfill
  \begin{subfigure}[t]{0.5\linewidth}
    \centering
    \includegraphics[width=\linewidth]{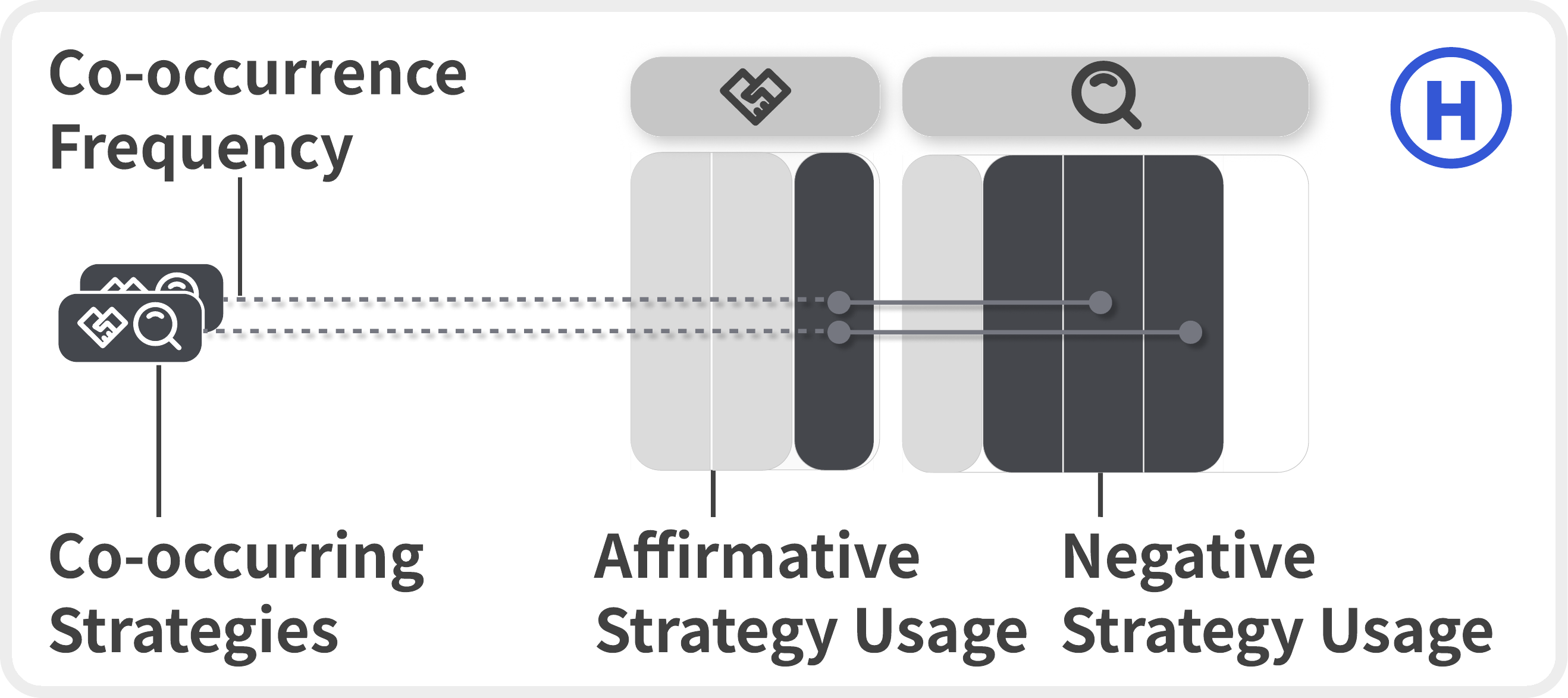}
    % \caption{}
    % \label{fig:sub2}
  \end{subfigure}
  \caption{The glyph design of the Strategy View. (G) augmented stacked bar design, where column height encodes session length and column width encodes the peak usage frequency of strategies within each session; (H) lines and boxes design, illustrating strategy co-occurrence frequency.}
  \label{fig:fig:strategy-def}
\end{figure}

The augmented stacked bar diagram is the core component of the \StgV{} (\cref{fig:teaser}A2), elaborately designed to display detailed strategy usage. 
This diagram is positioned alongside the \SeshV{} (\cref{fig:teaser}A0) with each of its rows directly corresponding to a circle in the \SeshV{} (\cref{fig:teaser}A0). 
The height of each row matches the circle's diameter, visually representing the length of that session's content. 
The diagram's columns represent different debate strategy types, and the width of each column is significant: it indicates the highest frequency (peak usage) of that particular strategy across all sessions (\cref{fig:fig:strategy-def}G). 
These columns are arranged from left to right, starting with the strategy that has the lowest peak usage (narrowest column) and progressing to the one with the highest (widest column). 
Within this structure, each small rectangle, signifies a single instance of a strategy being used. 
The color of the unit clearly indicates which debate side used the strategy, while its position in a column tells us which specific strategy it was; an icon at the top of each column further identifies the strategy type. 
To visualize co-occurrence, if a \bk{} involves multiple strategies, the units representing these strategies are connected by solid lines across their respective columns, as shown in \cref{fig:fig:strategy-def}H. 
The leftmost of these connected units then has a dashed line extending to a dedicated area on the diagram's left. 
This area displays icons representing all strategies that co-occurred in that specific block, with their colors also indicating the debate side. 
Furthermore, if the same set of co-occurring strategies appears multiple times within a session, their icon-based representations on the left may overlap. 
Each instance in the main bar chart will still link to this representation. 
Consequently, the number of dashed lines pointing to a particular box or overlapping area of co-occurring strategy icons directly reflects how often that specific combination of strategies was used together in the session.

\begin{figure}[htbp]
    \centering
    \begin{subfigure}[b]{0.32\linewidth}
        \centering
        \includegraphics[width=\linewidth]{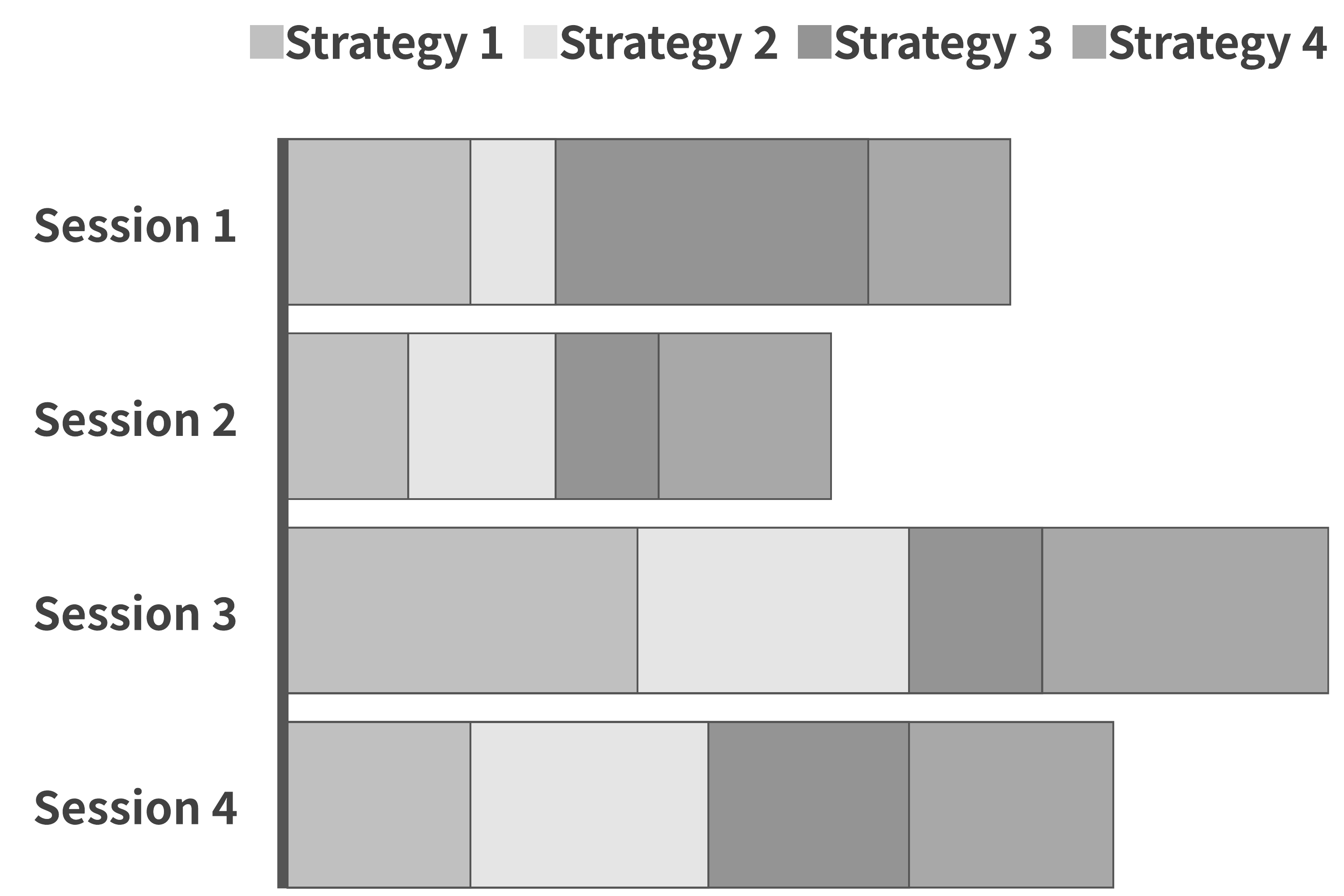}
        \caption{Stacked Bar Chart}
        \label{fig:group-contrast-sub1}
    \end{subfigure}
    \hfill
    \begin{subfigure}[b]{0.32\linewidth}
        \centering
        \includegraphics[width=\linewidth]{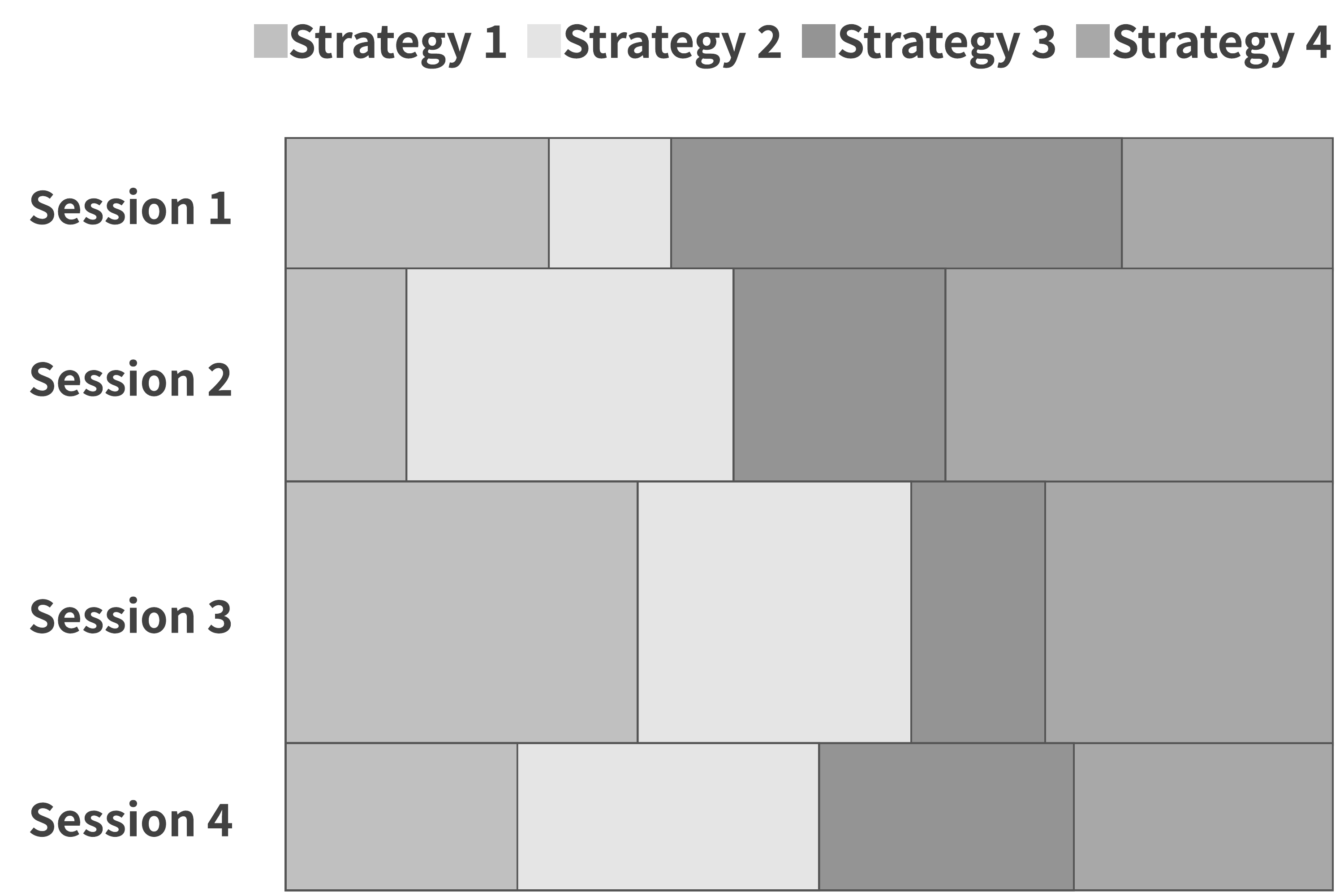}
        \caption{Mosaic Graph}
        \label{fig:group-contrast-sub2}
    \end{subfigure}
    \hfill
    \begin{subfigure}[b]{0.32\linewidth}
        \centering
        \includegraphics[width=\linewidth]{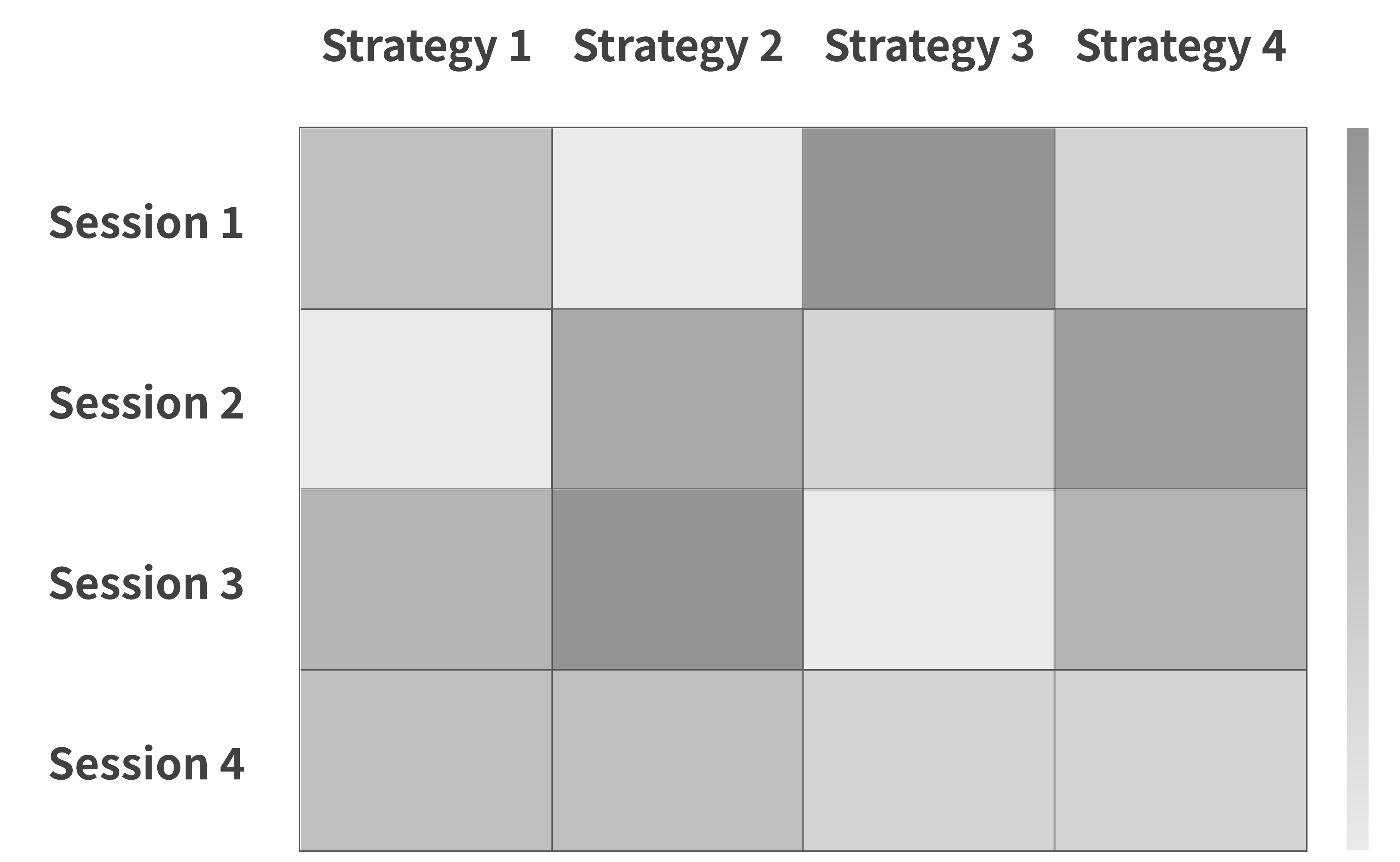}
        \caption{Heat Map}
        \label{fig:group-contrast-sub3}
    \end{subfigure}
    \caption{Alternative designs for the Strategy View.}
    \label{fig:group-contrast}
\end{figure}

\textbf{Design alternatives and Advantages.} 
Common visualization methods for this type of data, such as stacked bar charts, mosaic graphs, and heat maps, each present certain limitations when trying to clearly display all necessary information. 
Traditional stacked bar charts  (\cref{fig:group-contrast} a) make it difficult to accurately compare the sizes of different segments because they lack a common baseline, which also obscures the true differences between these segments. 
While mosaic graphs (\cref{fig:group-contrast} b) are more effective for showing proportions and the lengths of individual components, they still pose challenges for comparing components across different groups and tend to downplay the actual values involved. 
Similarly, heat maps (\cref{fig:group-contrast} c), though useful for illustrating the frequency of strategy use through color sequences, do not account for variations in debate duration, potentially leading to incorrect interpretations of the debate's overall structure. 
These limitations highlighted several key challenges for our design: accurately showing strategy usage across different sessions, enabling easy comparison of strategy use both within a single session and between multiple sessions, and clearly representing differences in content length from one session to another. 
To address these challenges, we developed an augmented stacked bar diagram. 
This new design builds upon the familiar concept of a stacked bar chart but introduces two significant enhancements: a variable-width for the bars and non-full-format filling. 
The primary advantage of our approach is its ability to integrate multiple dimensions of information — including the content length of each session, the specific strategies employed, and the distribution of these strategy usages across all sessions — into a single, comprehensive view.

\subsection{\CttV{}}
To help users examine the original text content (\textbf{R4}), the \CttV{} (\cref{fig:teaser}B) presents detailed content in a dedicated view. 
In this view, each \bk{} is shown as a card (\cref{fig:teaser}B1). 
Each card is color-coded by its side to indicate its origin. 
Users can select specific \bks{} to read their details together with the surrounding context.

On each card, \bk{} identifiers are in the upper-left corner, while debater identifiers (e.g., ``DEBATER A1'' means the affirmative side's first speaker) are placed in the upper-right corner. 
Below these identifiers, the card provides a collection of the \bk{}'s main \clashp{} and key viewpoints. 
This allows users to quickly grasp the main points before reading full texts.
To better show debate strategies within \bk{} content (\textbf{R2}), text in each \bk{} is segmented at sentences where strategies are applied. 
Strategy icons and descriptive labels are displayed beneath each text segment, indicating the specific approaches used.

\subsection{Interactions}
The three views in \toolName{} are interconnected with globally linked interactions. 
Selecting \bks{} highlights the corresponding spiral segments in the \PcsV{}, units in the \StgV{}, and cards in the \CttV{}, with the linked card auto-scrolling to the top. 
\toolName{} supports both session-based and \textbf{\textit{clash-point}}-based analysis alongside hovering interactions for condensed information display.

\textbf{Session-based Exploration.} 
Users are allowed to select a session by clicking a spiral segment or a circle in the \SeshV{}, activating the \CttV{} to display all \bks{} in the session and highlighting the corresponding row in the \StgV{}. 
Further selection of a turn via shorter spiral segments, or direct \bk{} selection via units in the \StgV{} or cards in the \CttV{} reveals contextual content and auto-scrolls the selected card.

\textbf{\textit{Clash-point}-based Exploration.} 
Clicking a \clashp{} in any view (legend, chordal graph, \PcsV{}, or \bk{} card) highlights relevant \bks{} and filters the display to its interaction chords. 
These chords are color-matched to the \clashp{}, showing the development of that specific disagreement (\textbf{R3}).

\textbf{Expandable Information Interaction.} 
Clicking on a condensed element, such as a disagreement block or a phrase in the \clashp{} legend, expands that legend to show more content. 
A further click on an individual disagreement in the legend reveals its full explanation. 
Hovering over various elements — including markers, icons, and phrases across the \clashp{}, \bk{}, and \StgV{} views — displays tooltips with their definitions.

\section{Case Study}
This section presents two case studies conducted on representative debate competitions in China and the United States to demonstrate the effectiveness of \toolName{}, by describing how two debate experts (\textbf{E1} and \textbf{E2}) used \toolName{} to explore and analyze debates.

\subsection{Case 1: Analyzing Debate Evolution and Interaction}

\begin{figure}[htbp]
    \centering
    \begin{subfigure}[b]{0.48\linewidth}
        \centering
        \includegraphics[width=\linewidth]{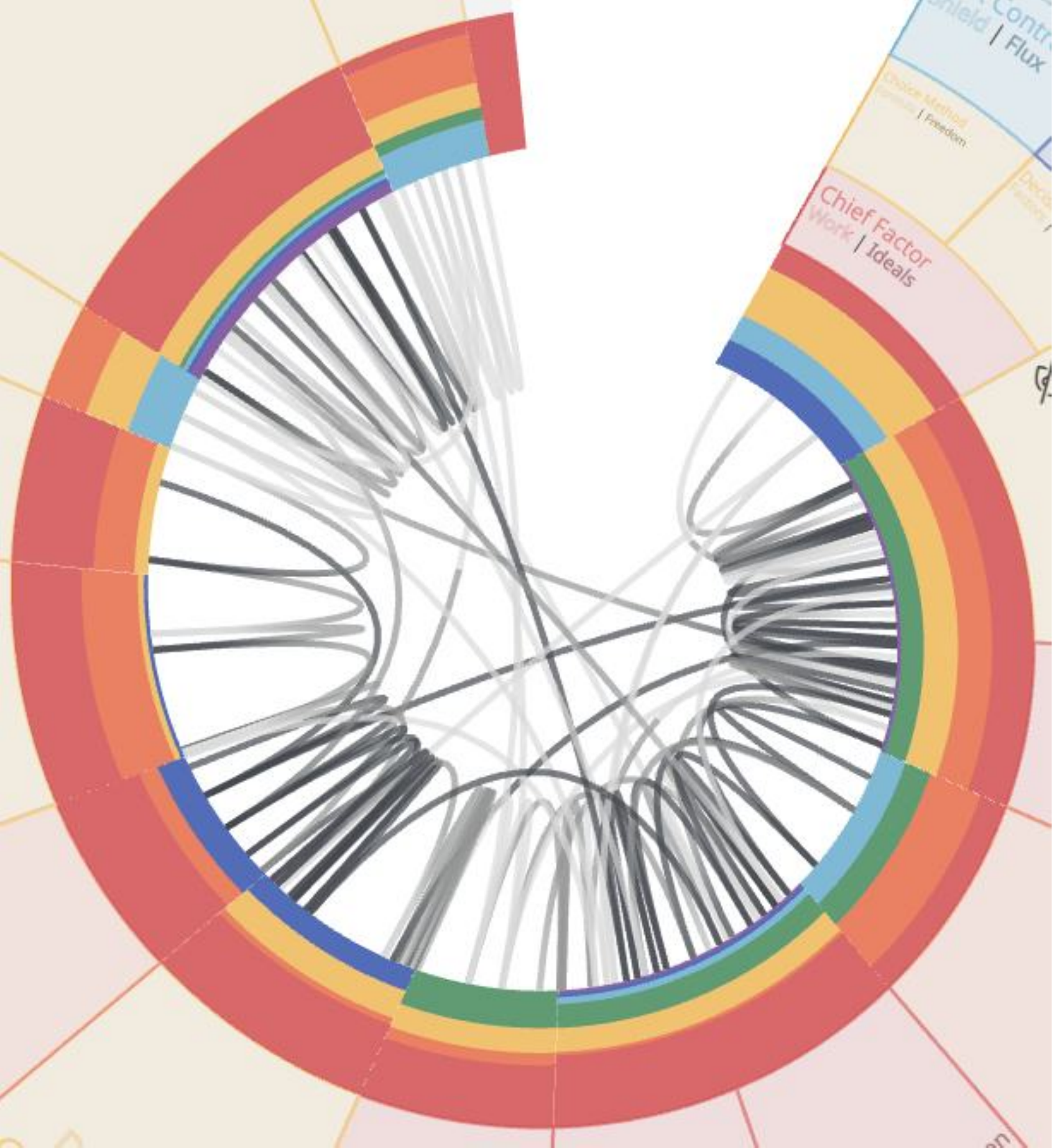}
        \caption{}
        \label{fig:c1-0}
    \end{subfigure}
    \hfill
    \begin{subfigure}[b]{0.48\linewidth}
        \centering
        \includegraphics[width=\linewidth]{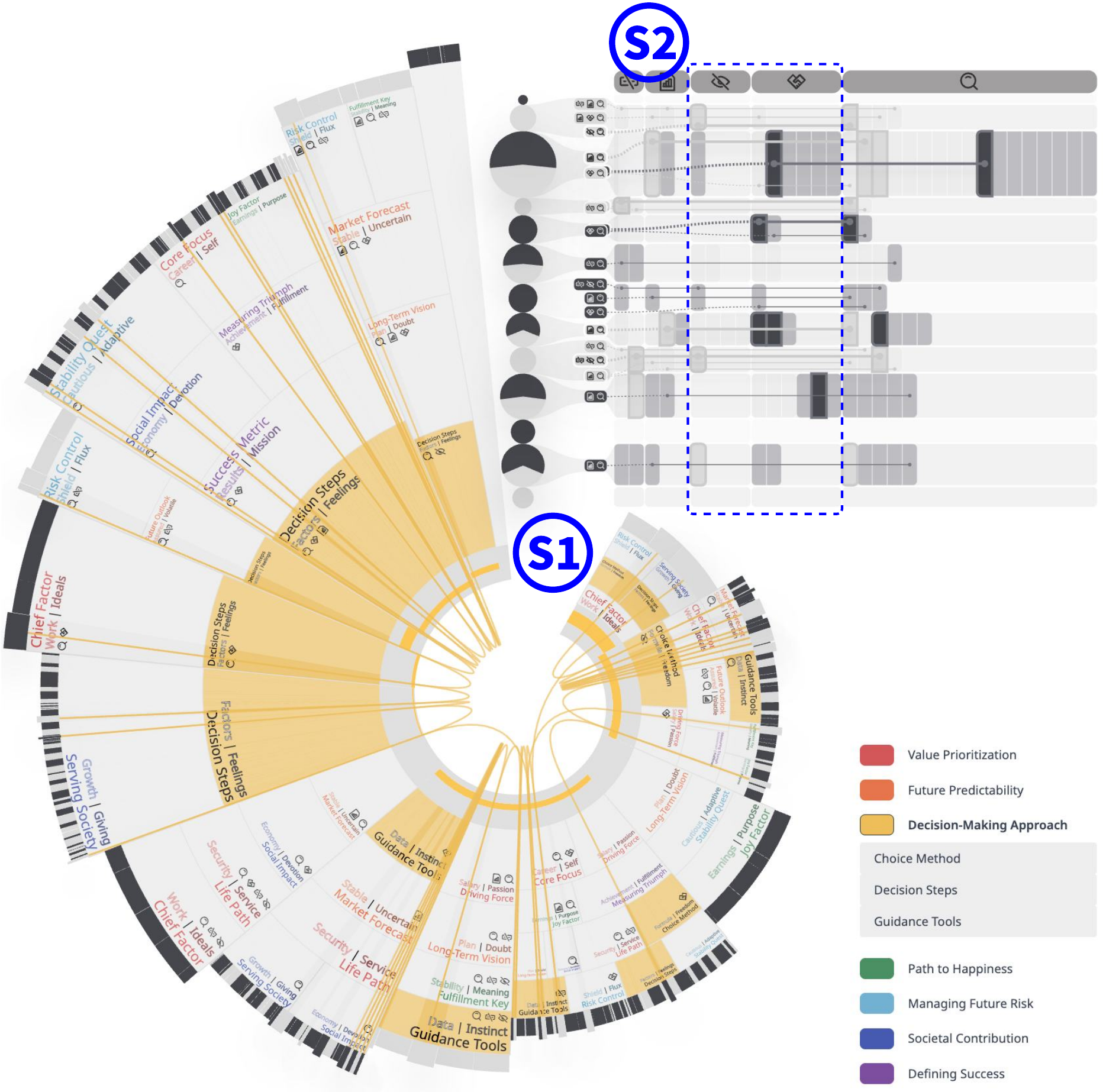}
        \caption{}
        \label{fig:c1-1}
    \end{subfigure}
    \caption{(a) Chordal diagram from Case Study 1. (b) S1: The disagreement and interaction path of the \clashp{} (``Decision-Making Approach''); S2: Strategies View from Case Study 1.}
    \label{fig:c1}
\end{figure}

\textbf{E1} is the captain of a Chinese debate competition team, who has five years of debate experience. She was asked to analyze a Chinese debate competition from ICDI by \toolName{}. The debate topic was ``Whether Employment Prospects Should Be the Primary Consideration When Choosing University Majors After the Gaokao\footnote{National College Entrance Examination (NCEE)}''. 
Firstly, she explored the Overview Part, as shown in \cref{fig:teaser}A. 
From the colorful filled blocks on the border of the chordal graph (\cref{fig:c1}(a)), she noticed that ``Value Prioritization'' was the main \clashp{} of the debate and was discussed in all sessions, as indicated by the continuous appearance and large proportion of the red blocks. 
In addition, she found ``Future Predictability'' and ``Decision-Making Approach'' were also discussed continuously, with orange and yellow blocks appearing in multiple session areas. However, ``Path to Happiness'' was mainly discussed in the first half of the debate, as green blocks mainly appeared in the earlier area. 
Moreover, blocks in other colors were found appearing scattered across sessions, and therefore considered the corresponding \clashp{} as non-core disagreements.

\begin{figure}[htbp]
    \centering
    \begin{subfigure}[b]{0.2\linewidth}
        \centering
        \includegraphics[width=\linewidth]{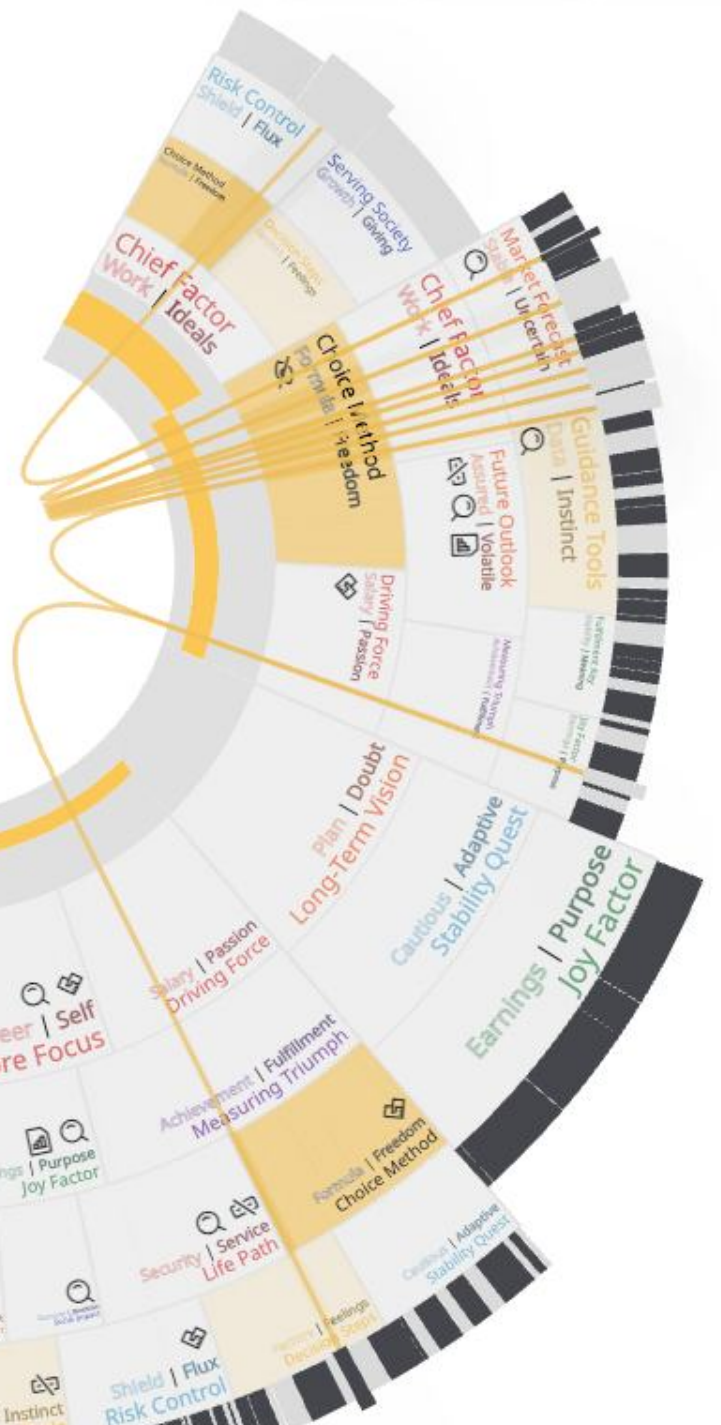}
        \caption{}
        \label{fig:c1-CM}
    \end{subfigure}
    \hfill
    \begin{subfigure}[b]{0.47\linewidth}
        \centering
        \includegraphics[width=\linewidth]{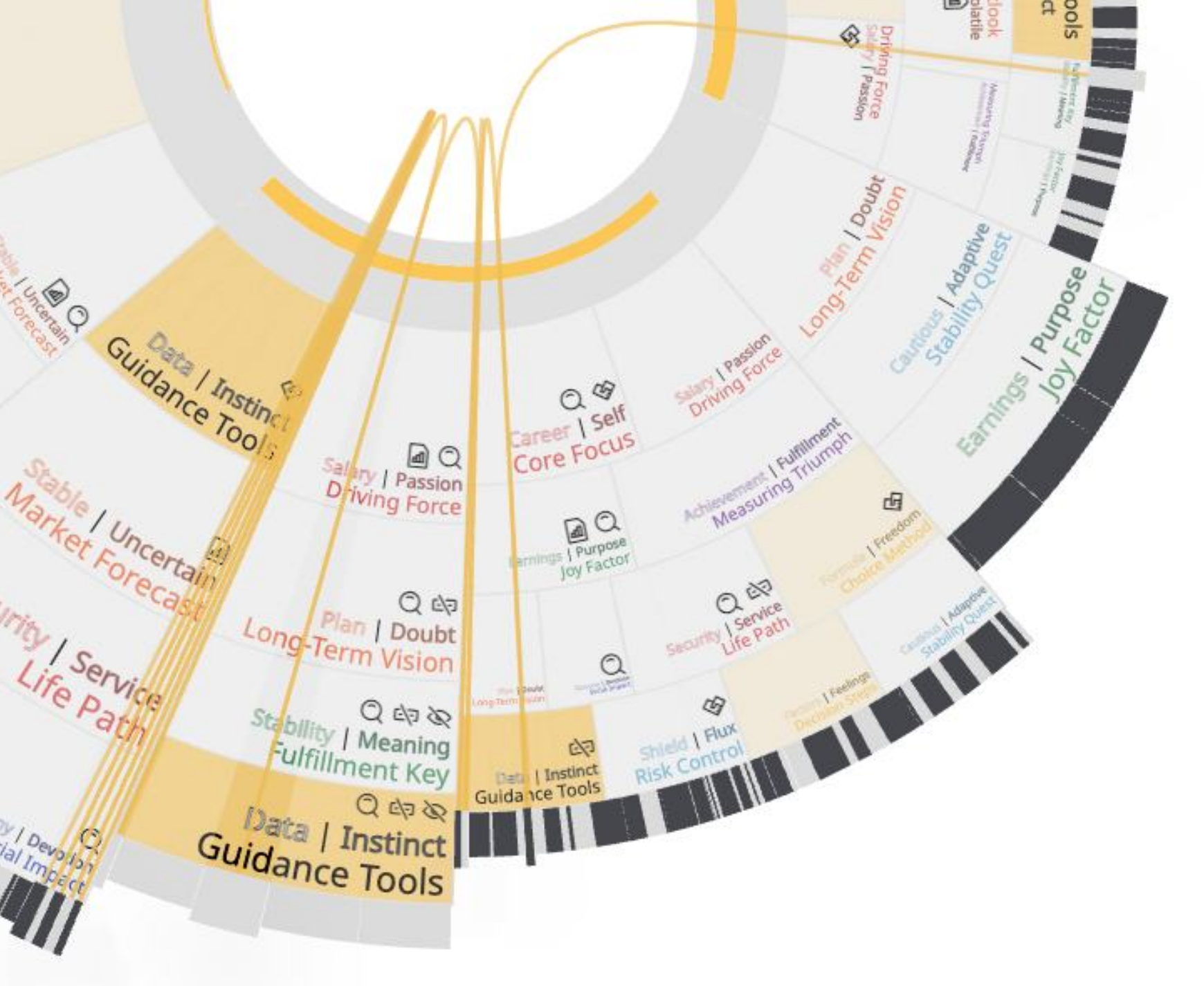}
        \caption{}
        \label{fig:c1-GT}
    \end{subfigure}
    \hfill
    \begin{subfigure}[b]{0.31\linewidth}
        \centering
        \includegraphics[width=\linewidth]{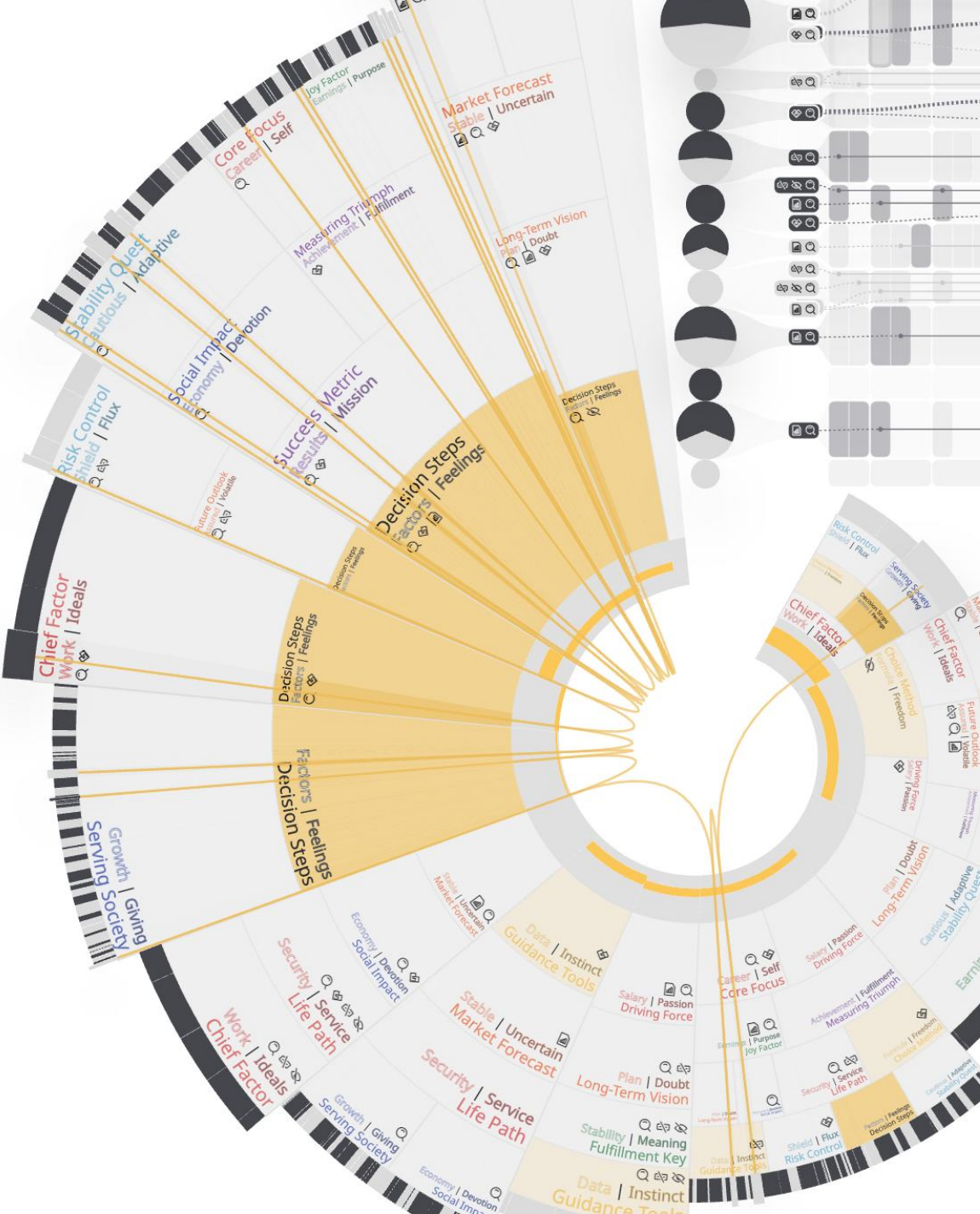}
        \caption{}
        \label{fig:c1-DS}
    \end{subfigure}
    \caption{Interaction paths within the \clashp{} (``Decision-Making Approach''): (a) ``Choice Method''. (b) ``Guidance Tool''. (c) ``Decision Steps''.}
    \label{fig:c1-DMA}
\end{figure}

\textbf{Explore the \clashps{} and the strategies.} \textbf{E1} noticed that yellow blocks appeared only in sessions involving the affirmative side (S1 in \cref{fig:c1}(b)). 
Therefore, she started focusing on the corresponding \clashp{} — ``Decision-Making Approach''. 
Based on this observation, she concluded that the affirmative side had an advantage on this \clashp{}, as the negative side rarely participated in related discussions. 
Then, she used the strategy view (S2 in \cref{fig:c1}(b)) to explore the strategies used in relevant sessions. 
She noticed that black \bks{}, representing the negative side, appeared in the column marked with a handshake icon (``Refutation through Agreement''), but never appeared in the column marked with an eye icon (``Refutation through Ignoring''). 
This observation further confirmed her judgment, as the negative's strategy was mainly to agree with the affirmative side.
Consequently, she further analyzed the disagreements and related interactions of this \clashp{}. 
She chose the three disagreements of this \clashp{} in turn and observed their development path in the chordal graph. 
As shown in \cref{fig:c1-CM}, \cref{fig:c1-GT} and \cref{fig:c1-DS}, \textbf{E1} found the sequential clustering of yellow blocks, 
and these three disagreements were all initiated by the affirmative side, clearly evolved one after another over time. 
Specifically, the disagreement ``Choice Method'' appeared first, starting from session 1 and lasting for four sessions, which formed the earliest cluster of yellow blocks (\cref{fig:c1-CM}). 
Immediately afterward, the disagreement ``Guidance Tool'' became the main topic and was actively discussed until session 5, forming the second clear cluster (\cref{fig:c1-GT}). 
Finally, in the second half of the debate, the disagreement ``Decision Steps'' emerged as the primary focus, as it was heavily discussed during the final six sessions and formed the last cluster of yellow blocks (\cref{fig:c1-DS}). 

\textbf{Delve into the interaction within a disagreement.} 
By reviewing the three progressively developed disagreements and related viewpoints, \textbf{E1} found the negative side hold the view that ``people should first reflect on themselves to learn what they want to learn before choosing college majors''.
Since ``Decision Steps'' was the last disagreement of this \clashp{}, whoever had an advantage on this disagreement would determine who won the entire \clashp{}. 
E1 then followed the path under ``Decision Steps'' (\cref{fig:c1-DS}) to examine the detailed content. 
She noticed that both the beginning and end of the path corresponded to white blocks, and a long black block existed in the middle of the path. 
She focused particularly on understanding these three corresponding blocks. 
Through careful examination, she found that the affirmative side presented this disagreement using detailed and solid data, while the negative side only briefly argued their position and explained their position by ``ideal''. 
In the final block, the affirmative side pointed out the negative side's flaw of inconsistent reasoning, criticizing them for shifting from ``reflecting on oneself'' to ``idealism'' above all. 
Ultimately, she concluded that the affirmative side won this \clashp{}. 
Interestingly, she noted that the negative side's initial suggestion of ``reflecting on oneself'' did not even serve as a valid rebuttal to the affirmative side. 
If a person is indeed materialistic, then ``reflecting on oneself'' would naturally lead them to choose their major based on employment opportunities, which clearly aligns with the affirmative side's position. 
This suggests that the negative side itself has a flaw in its logic.

\subsection{Case 2: Understanding Overall Debate tactics}

\textbf{E2} is the captain of an English debate club, who primarily coaches beginners. 
She was asked to analyze an English competition of NSDT by \toolName{}, with the topic ``This House Believes That Housing Is a Guaranteed Right''. 
\textbf{E2} explained to us that, compared to being a debater who deeply explores every detail during a debate, she — as a coach — pays more attention to the high-level, overall tactics of the debate. 

\begin{figure}
    \centering
    \includegraphics[width=\linewidth]{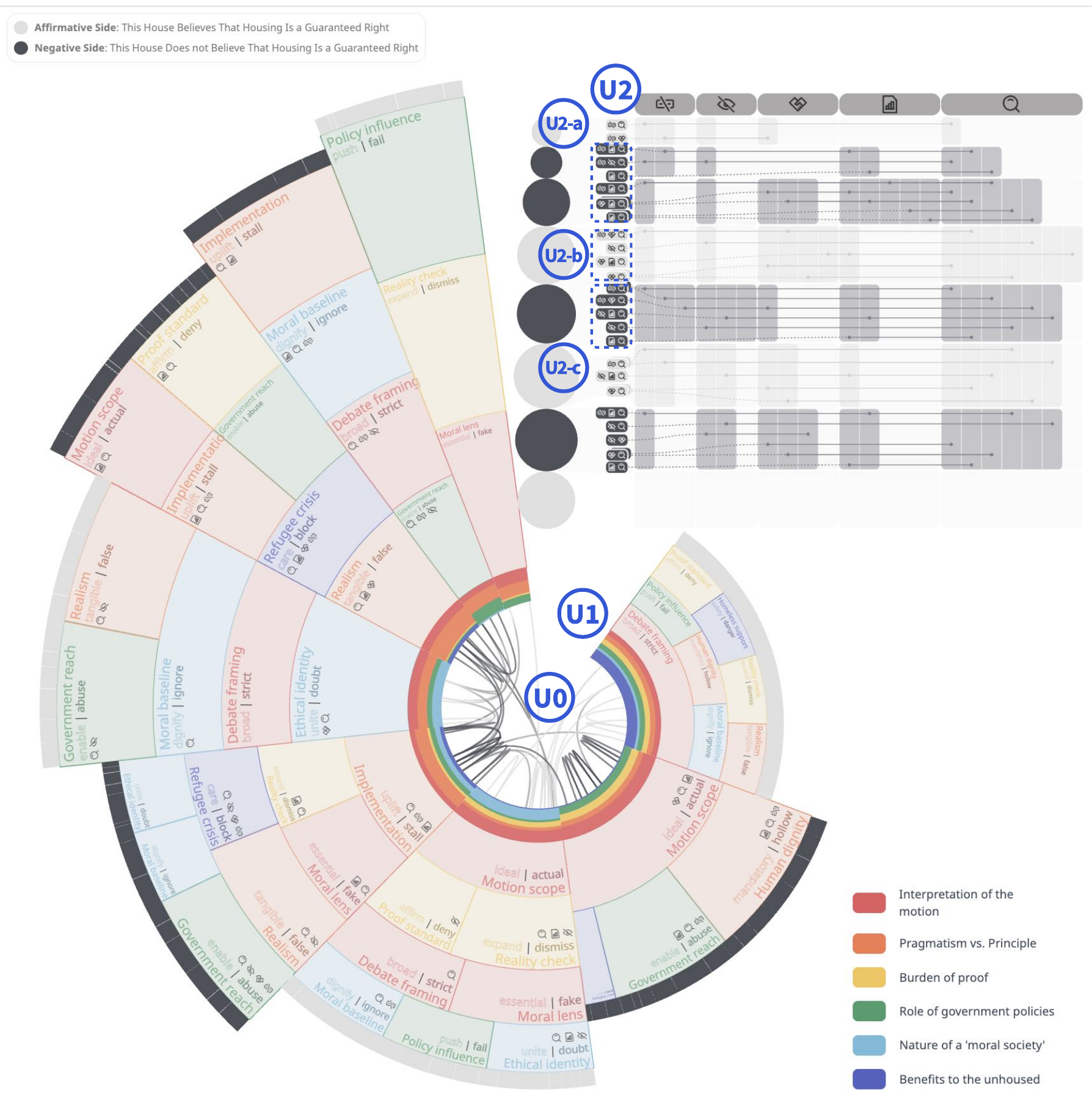}
    \caption{The chord diagram (U0) shows few interactions between the two sides, indicated by black-and-white lines. The \clashp{} ``Interpretation of the motion'' (U1, red) reveals fundamental differences in how each side defined the debate topic. Additional \clashps{}, ``Pragmatism vs. Principle'' (orange) and ``Role of government policies'' (yellow), illustrate independent argumentation from each side. The \StgV{} (U2) highlights clear differences in strategy patterns between the two sides.}
    \label{fig:c2}
\end{figure}

\textbf{E2} first noticed that black and white chords appeared separately in the chord diagram, with very few chords mixing black and white (\cref{fig:c2}U0). 
This indicated that the affirmative and negative sides mainly developed arguments internally within their own sides, rather than between the two sides, which means direct interactions and clashes between the two sides are limited. 
She found this phenomenon unusual. 
Therefore, she further looked into the \clashp{} labeled ``Interpretation of the motion'', represented by the red color sections that frequently appeared and occupied significant space (\cref{fig:c2}U1). 
She discovered that the two sides differed fundamentally in their understanding and definitions of the debate topic itself. 
This difference explained why there were relatively few direct confrontations between them.

Next, \textbf{E2} examined two other \clashps{}, ``Pragmatism vs. Principle'' and ``Role of government policies'', represented by orange and yellow colors respectively, as these also occupied large spaces. 
She found that under these two \clashps{}, the two sides had different definitions of ``should'', and both sides continued independently arguing their own interpretations of the motion: the affirmative side argued that ``Housing ought to be a guaranteed right'', while the negative side argued that ``Housing is already a guaranteed right''.

Given these observations, \textbf{E2} shifted her attention to the strategy view (\cref{fig:c2}U2) to better understand the strategic framework of the entire debate. 
She noticed that black blocks appeared extensively across all columns, whereas white blocks were more concentrated in the columns labeled ``Refutation through Agreement'' and ``Refutation through Reasoning''. 
Considering both sides' differing definitions of the motion, she identified distinctly opposite patterns between the two teams: the affirmative side primarily used reasoning to indirectly strengthen their own definition, rarely directly refuting the negative side. 
In contrast, the negative side heavily relied on evidence to directly refute the affirmative side.

\textbf{E2} further examined co-occurring strategies and discovered that for the black side (negative), ``Refutation through Reasoning'' frequently appeared together with ``Refutation through Evidence'' (\cref{fig:c2}U2-a, \cref{fig:c2}U2-c). 
On the other hand, for the white side (affirmative), ``Refutation through Reasoning'' was more often combined with ``Refutation through Agreement'' (\cref{fig:c2}U2-b). 
She concluded that these insights further highlighted the systematic differences between the affirmative side and negative side throughout the debate, not only on basic interpretations, but also on strategic framework.

\section{User Study}

\begin{figure}[htbp]
    \centering
    \includegraphics[width=\linewidth]{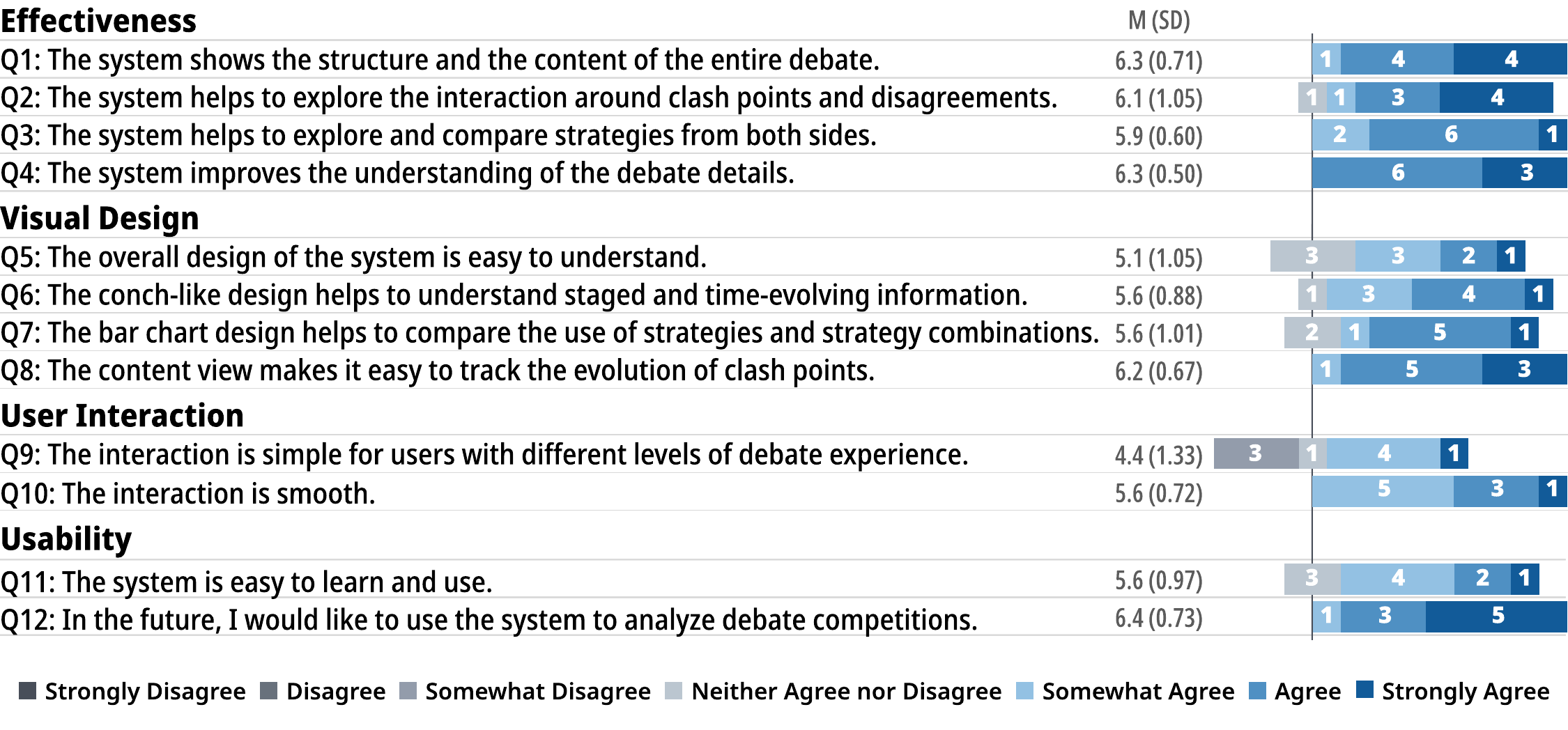}
    \caption{User interview questionnaire results, including the effectiveness, visual design, interaction, and usability, rated on a 7-point Likert scale.}
    \label{fig:user-score}
\end{figure}

We conducted a user study by comparing the effects of debate analysis using \toolName{} and other common analyzing methods: watching debate videos and reading debate transcripts. 
Manual debate analysis is time-consuming and cognitively demanding \cite{3Supardi_Sayogie_2022} because it is burdensome for humans to identify and piece together the logic of \clashps{} and refutation strategies from the extensive context of the entire debate. Therefore, when analyzing a debate, human debaters or coaches will inevitably take longer time and confront growing cognitive load. 
Taking the above into consideration, we measured the time participants spent to evaluate efficiency, their cognitive load during analysis to assess the effectiveness \cite{Leppink_2013}, and their feedback and ratings for system to understand user satisfaction and perceived usability.

\subsection{Study Design}

\textbf{Group design.}
In this study, we aimed to evaluate whether the use of \toolName{} improves the debate analysis effects. 
We conducted a controlled experiment consisting of one experimental group and two control groups. 
Participants from all three groups were asked to analyze the same debate competition through a series of tasks. 
Specifically, participants in the experimental group were asked to use \toolName{} to analyze the debate, while participants in the video control group and the text control group analyzed the debate using video and text respectively. 

\textbf{Data and tasks.} 
The video and text data from the ICDI were used as the experimental material, as detailed in \cref{tab:dataset}. 
This competition was chosen because it contains richer \clashps{} and disagreements. 
Moreover, since the data is in Chinese, it reduces participants' cross-language comprehension burden. 
We designed seven debate analysis tasks (\cref{tab:tasks}) to measure the completion rate of debate analysis from three aspects: what to debate (T1-T3), how to debate (T4-T5) and comprehensive analysis (T6-T7). 
Following these tasks, we assessed participants' cognitive load through a ten-item questionnaire for the measurement of \textbf{IL} (\textbf{I}ntrinsic \textbf{L}oad), \textbf{EL} (\textbf{E}xtraneous \textbf{L}oad),and \textbf{GL} (\textbf{G}ermane \textbf{L}oad) \cite{Leppink_2013}. 
The detailed content of this questionnaire can be found in the supplementary materials. 
Additionally, a Likert scale (1 = strongly disagree, 7 = strongly agree) was designed for participants in the experimental group to evaluate the effectiveness, visual design, interaction and usability of \toolName{} (Q1 - Q12 in \cref{fig:user-score}).

\begin{table}[htbp]
\centering
\caption{The tasks for participants to perform in our user interviews.}
\label{tab:tasks}
\resizebox{\columnwidth}{!}{%
\begin{tabular}{ll}
\toprule
\multicolumn{2}{l}{\textit{A. What to Debate}} \\
T1 & Summarize how the content of both sides evolved across different sessions. \\
T2 & List the clash points in this debate. \\
T3 & Review how debaters had discussions around clash points and list the most impressive interactions. \\
\midrule
\multicolumn{2}{l}{\textit{B. How to Debate}} \\
T4 & List strategies used by debaters when making rebuttals. \\
T5 & List strategies that were commonly combined and used together. \\
\midrule
\multicolumn{2}{l}{\textit{C. Comprehensive Analysis}} \\
T6 & Summarize the performance of both sides. \\
T7 & Summarize what you have learned. \\
\bottomrule
\end{tabular}%
}
\end{table}

\textbf{Participants.} 
We recruited 27 debaters and debate coaches to participate in the user study (7 female, 20 male; age: $ \text{Mean} (M) = 21.0$, $ \text{Standard Deviation} (SD) = 2.4$). 
All of them have 1-9 years of related experience in debate and debate analysis. 
Specifically, 19 of them have 1-3 years' experience, 4 have 3-5 years' experience and 4 have more than 5 years' experience. 
Among these participants, two (E3, E4) are world champions in Chinese debate competitions. 
Based on their length of debate experience, we randomly divided them into three groups through stratified sampling, ensuring similar experience levels within each group. 
Additionally, another expert (E5), who is also a world champion and the ``Best Debater'' in Chinese debate competitions, provided feedback on our system.

\textbf{Procedure.} 
The study was conducted both online and offline. 
Participants could choose their types of attendance based on their location and convenience.
The \toolName{} system, the debate video and the debate text were all accessible online. 
We first obtained their consent, and introduced the research background and experiment process for approximately 5 minutes. 
For the experimental group, we additionally provided a 15-minute introduction to the functions of \toolName{}. Participants in this group were then allowed to freely explore the system until they were familiar with how to use it. 
After the introduction, participants in three groups should complete seven tasks by using \toolName{}, watching debate video and reading debate text, respectively. 
Then, participants completed the ten-item cognitive load questionnaire, and participants in the experimental group also filled a system evaluation questionnaire. 
Task completion times spanned between 30-135 minutes ($M = 73.9$, $SD = 26.9$), with a debate duration of 66 minutes.

\subsection{Results}

\begin{figure*}[htbp]
    \centering
    \begin{subfigure}[b]{0.3\linewidth}
        \centering
        \includegraphics[width=\linewidth]{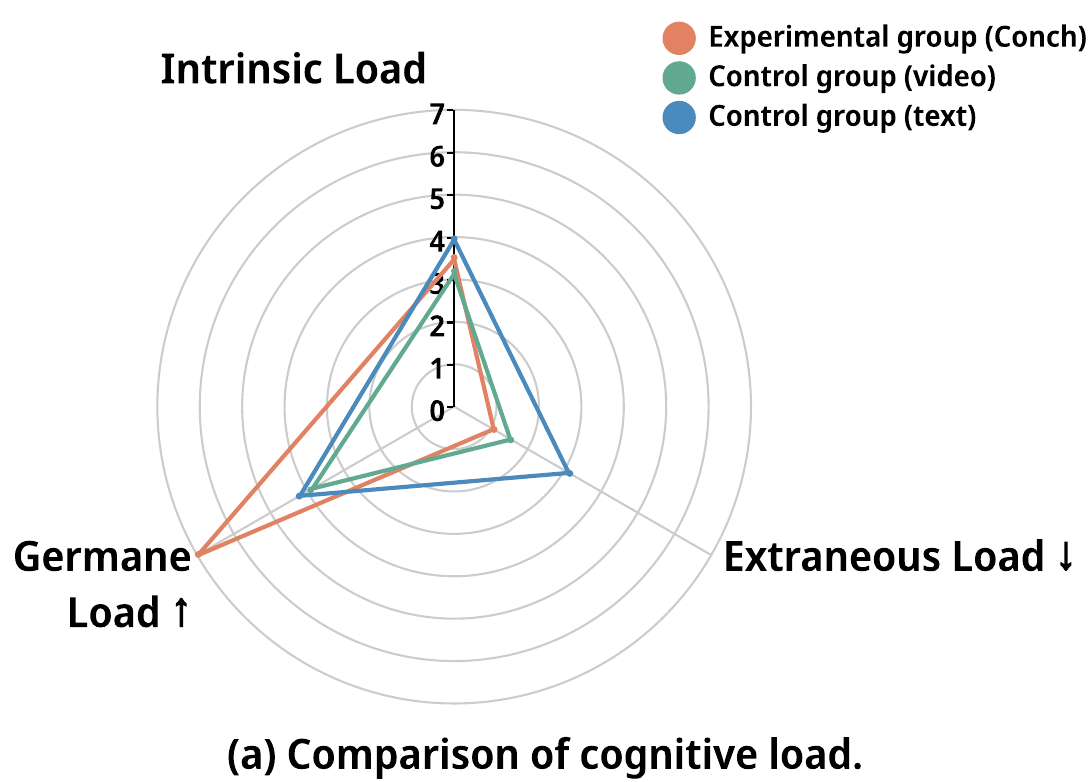}
    \end{subfigure}
    \hfill
    \begin{subfigure}[b]{0.68\linewidth}
        \centering
        \includegraphics[width=\linewidth]{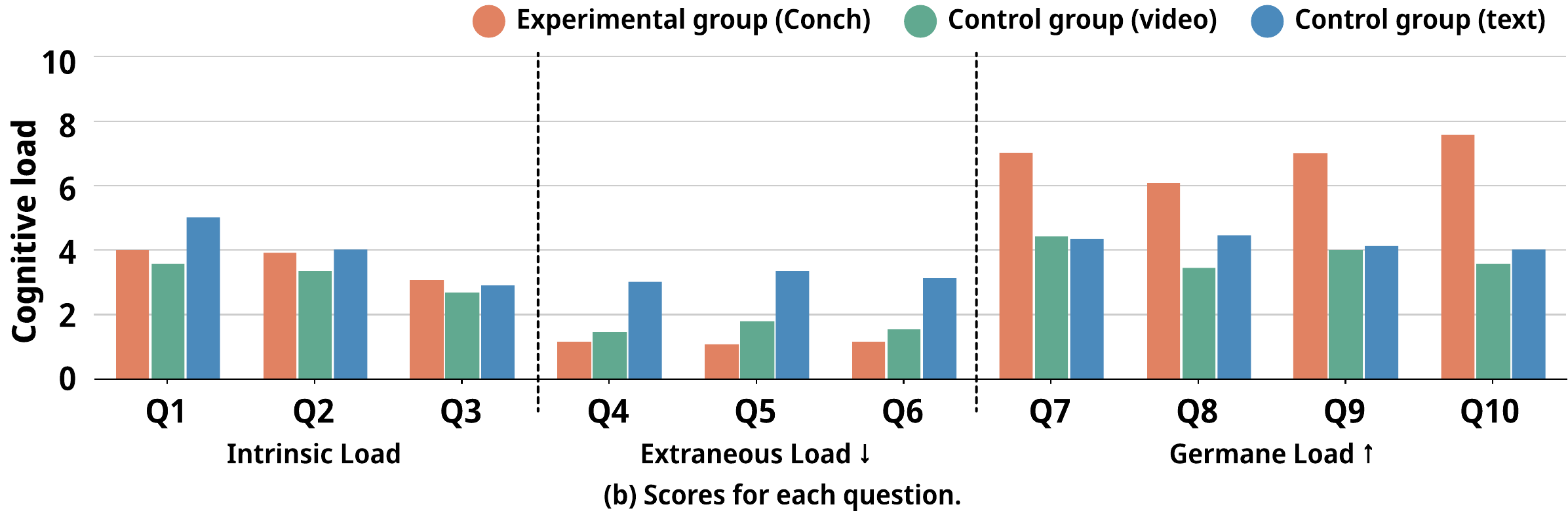}
    \end{subfigure}
    \caption{Cognitive load and performance comparison among the experimental and control groups. (a) Comparison of intrinsic, extraneous, and germane cognitive load among the experimental and control groups. (b) Scores for each question across the experimental and control groups.}
    \label{fig:cl}
\end{figure*}

Upon completing the debate analysis task, the experimental group using \toolName{} showed shorter analyzing time than other two control groups and a better cognitive load profile for learning. 

\textbf{Task Completion Rate.} 
All participants successfully completed the entire debate analysis task. 
We found that the experimental group (using \toolName{}) and the two control groups all achieved a 100\% completion rate. 
This indicates that \toolName{} is as effective as the baseline methods in enabling users to finish the required analysis, ensuring that the tool does not introduce barriers to task completion.

\begin{figure}[htbp]
    \centering
    \includegraphics[width=\linewidth]{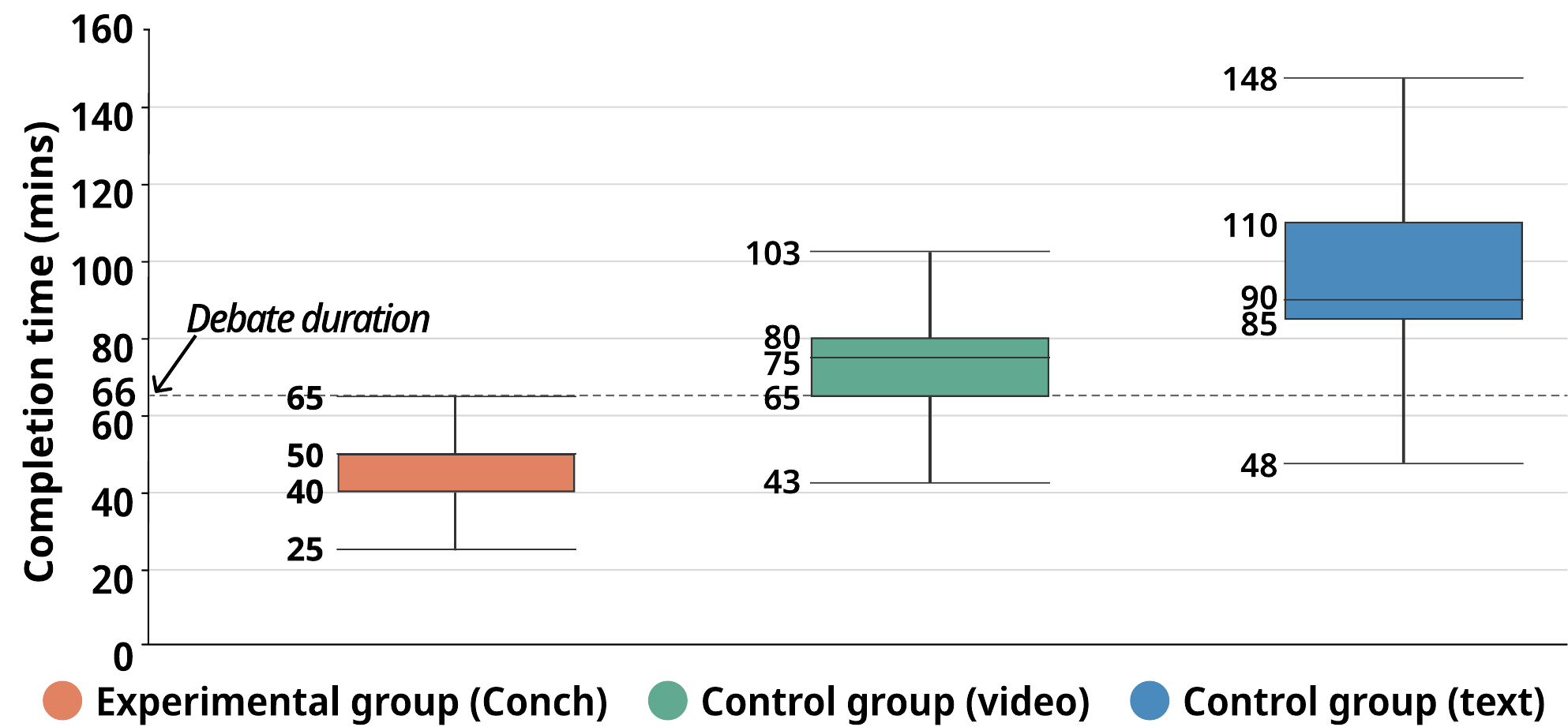}
    \caption{Total task completion time for the experimental group and two control groups.}
    \label{fig:time-spend}
\end{figure}

\textbf{Completion time.} 
As shown in \cref{fig:time-spend}, participants in the experimental group (A) completed the tasks faster than those in the video control group (B) and the text control group (C). 
Specifically, participants using \toolName{} in the experimental group ($M_A = 51.1$, $SD_A = 20.2$) spent 32.8\% less time than participants watching debate video ($M_B = 76.1$, $SD_B = 17.8$) and 45.9\% less than those reading debate text ($M_C = 94.4$, $SD_C = 23.8$), which demonstrated statistically significant differences ($p_B < 0.01$, $p_C < 0.01$) suggested by the Mann-Whitney U test \cite{mwut}. 
Since debate analysis requires a global consideration of the entire content, single task is difficult to be completed and timed independently.

\textbf{Cognitive Load.} 
Cognitive load consists of intrinsic load (IL), extraneous load (EL), and germane load (GL). IL relates to task complexity and participants' prior knowledge, EL arises from unhelpful instructional features, and GL results from beneficial instructional features \cite{Leppink_2013}. 
Thus, lower EL and higher GL indicate whether a system provides better support for users in performing tasks effectively.

As shown in \cref{fig:cl}, we found that IL scores were similar across all three groups (experimental: $3.67\pm2.17$, control-video: $3.26\pm1.73$, control-text: $3.96\pm2.02$), indicating comparable task difficulty and participant knowledge. 
Using the Mann-Whitney U test, the experimental group had significantly lower EL than the text control group ($1.15\pm1.12$ \textit{vs.} $3.15\pm2.33$, $U=17.5$, $p=0.046$). 
Compared to the video control group, the experimental group's EL was lower on average, but the difference was not statistically significant ($1.15\pm1.12$ \textit{vs.} $1.59\pm2.34$, $U=44.5$, $p=0.755$). 
Additionally, the experimental group's GL was significantly higher compared to both the video-based group ($7.03\pm1.85$ \textit{vs.} $4.22\pm2.58$, $U=66.5$, $p=0.024$) and the text-based group ($7.03\pm1.85$ \textit{vs.} $3.94\pm2.58$, $U=64.0$, $p=0.042$).

\textbf{Questionnaire.} 
\cref{fig:user-score} reports the results of the user interview questionnaire across four dimensions: effectiveness, visual design, user interaction, and usability. 
Overall, participants in the experimental group gave high ratings to \toolName{}, with most of the closed-ended questions receiving positive feedback.  
Specifically, participants expressed satisfaction with the effectiveness and usability of the system, as they rated questions in these dimensions around 6 out of 7. 
They also provided generally positive ratings for visual design and user interaction, although scores in these two dimensions were slightly less consistent. 
However, scores for Q5 (``easy to understand'') and Q9 (``interaction is easy for users with any level of debate experience'') were relatively lower. 
For Q5, participants indicated that the visual design was somewhat challenging to understand due to the introduction of multiple visual components and novel visualization methods. 
Regarding Q9, participants indicated that the system offered different benefits based on user experience. 
For beginners, \toolName{} served as a guide to understand the core elements of a debate, such as \clashps{} and strategies; for experts, it functioned as a tool to make their analysis more efficient.

In the open-ended feedback, participants suggested that \toolName{} was broadly beneficial. 
For example, experts E3 and E4 noted that the \PcsV{} was effective for tracking the evolution of arguments, showing which points were introduced, dropped, or modified over time. 
E5 added that the \PcsV{} (\cref{fig:teaser}A1) clearly highlighted the main conflicts. 
The \StgV{} (\cref{fig:teaser}A2) was also widely praised for summarizing common tactics and identifying strategic strengths or weaknesses. 
Finally, E5 found the \CttV{} (\cref{fig:teaser}B) useful for filtering information to focus on the most meaningful content.
Overall, most of users reported that \toolName{} effectively helped them quickly learn main arguments, different strategies, and the clear evolution of debates.

\textbf{Limitations and improvements.} 
First, \toolName{} doesn't show the performance of each single debater. 
This results from our system's main focus: the presentation of logical and structural debate content, rather than individual debater's performance. 
In future work, we can develop a debater's view to support speaker-level analysis, including the duration of participation, the number of contributions made, and the efficiency of argumentative output. 
This would enable more fine-grained assessments of participant performance. 
Second, current LLMs are still constrained by context window size and reasoning capabilities, which can lead to mistakes in the structured output, causing confusion in user interpretation. 
In future work, it is worthwhile to fine-tune a LLM with a long context window and strong reasoning abilities to generate all structured analysis results in a single pass.

\section{Discussion}
In this section, we summarize the lessons we learned during the development and evaluation of \toolName{}.

\textbf{Transferability of structured text analysis.} 
We used the competitive debate to characterize the problem domain. 
The workflow can also be applied to legal text interpretation \cite{caselaw2024}, as shown in \cref{fig:tech-flow}. 
This is because structured texts are characterized by logical frameworks, argumentation, and organized reasoning. 
We also captured the strategies and patterns used by speakers during argumentation. 
This enriched our understanding of content interaction and further supported the transferability of our approach.

\textbf{Enhancing sequential textual analysis through hierarchical segmentation.}  
Sequential textual data often lacks clear structure because content is presented incrementally over time, making it challenging to capture relationships and track evolving contexts effectively.
While overly general summaries tend to omit essential details critical for user judgment, excessively detailed representations of key points can overwhelm users and obscure content relationships. 
We propose a hierarchical segmentation approach to balance macro-level summaries and detailed key points. Specifically, our method integrates high-level key points (e.g., clash points) with intermediate-level structural anchors (e.g., disagreements).
This hierarchical segmentation framework should be used to enhance the structural organization of sequential textual data. 
It could facilitate clearer content analysis and improve user comprehension in scenarios such as online discussions, educational forums, and collaborative reviews.

\section{Conclusion}
In this paper, we presented \toolName{}, an interactive visualization system designed to assist debaters and coaches in analyzing \cds{}. 
In our system, LLM-based natural language processing techniques with carefully designed prompts are utilized to structure debate content and identify multi-level semantic information, including clash points, disagreements, viewpoints, and argumentative paths, enabling comprehensive content analysis. 
The novel visualization designs, including concentric spirals and augmented stacked bar charts, effectively illustrate the evolution of debate content and the strategies adopted by debaters.
Through two case studies and a carefully-designed user study, we demonstrated \toolName{}'s effectiveness in helping users gain deep insights into various aspects of debate competitions, including content evolution, strategy usage, and interactions. 

In future work, we will extend \toolName{} to analyze everyday communication, which lacks the clear logical rules of structured arguments. 
We will also incorporate emotion analysis to investigate how emotional factors in these texts influence a reader's acceptance and agreement.

\section*{Acknowledgments}
This work is supported in part by the Taihu Lake Innovation Fund for Future Technology, 
Huazhong University of Science and Technology (HUST), 
under Grant 2023-B-8; 
in part by the Fundamental Research Funds for the Central Universities, HUST, 
under Grant 82400049. 
The computation was completed in the HPC platform of Huazhong University of Science and Technology. 
Ran Wang is the corresponding author (rex\_wang@hust.edu.cn).

\bibliographystyle{abbrv-doi-hyperref}

\bibliography{template}

\appendix % You can use the `hideappendix` class option to skip everything after \appendix

\end{document}